\documentclass{aa}

\usepackage{graphicx}
\usepackage{txfonts}
\usepackage{caption}
\usepackage{placeins}
\usepackage{amsmath}    
\usepackage{amssymb}
\usepackage{threeparttablex}
\usepackage{hyperref}
\usepackage{url}
\usepackage[dvipsnames]{xcolor}
\hypersetup{
  colorlinks   = true, 
  urlcolor     = blue, 
  linkcolor    = blue, 
  citecolor   = blue 
}
\usepackage{natbib}
\usepackage{orcidlink}
\bibpunct{(}{)}{;}{a}{}{,}

\begin{document}

   \title{Evidence of a gap in the envelope mass fraction of sub-Saturns}
   
   \author{Luis Thomas\thanks{Email:lthomas@mpe.mpg.de}\orcidlink{0009-0006-1571-0306}\inst{1,2}
          \and
          Louise D. Nielsen \orcidlink{0000-0002-5254-2499} \inst{1}
          \and
          Lorena Acuña-Aguirre \orcidlink{0000-0002-9147-7925} \inst{3}
          \and
          Alex Cridland \orcidlink{0000-0003-1777-9930} \inst{1}
}

   \institute{
   University Observatory Munich, Faculty of Physics, Ludwig-Maximilians-Universit\"at München, Scheinerstr. 1, 81679 Munich, Germany
    \and
    Max-Planck Institute for Extraterrestrial Physics, Giessenbachstrasse 1, D-85748 Garching, Germany
    \and 
    Max Planck Institut für Astronomie, Königstuhl 17, 69117 Heidelberg, Germany
    }
   \date{}

  \abstract 
  {
  Under the core-accretion model, gas giants form via runaway accretion. This process starts when the mass of the accreted envelope becomes equal to the mass of the core. We modeled a population of warm sub-Saturns to search for imprints of their formation history in their internal structure. Using the GAS gianT modeL for Interiors (GASTLI), we calculated a grid of interior structure models on which we performed retrievals for our sample of 28 sub-Saturns to derive their envelope mass fractions ($f_{env}$). For each planet, we ran three different retrievals, assuming low ($ -2.0 < \log(\mathrm{Fe}/\mathrm{H}) < 0.5$), medium ($ 0.5 < \log(\mathrm{Fe}/\mathrm{H}) < 1.4$), and high ($ 1.4 < \log(\mathrm{Fe}/\mathrm{H}) < 1.7$) atmospheric metallicity. 
 The distribution of $f_{env}$ in our sample was then compared to outcomes and predictions of planet formation models. When our results are compared to the outcomes of a planetesimal accretion formation model, we find that we require a high atmospheric metallicity for intermediate-mass sub-Saturns to reproduce the simulated planet population. For higher planetary masses, a medium atmospheric metallicity provides the best agreement. Additionally, we find a bimodal distribution of $f_{env}$ in our sample with a gap that is located at different values of $f_{env}$ for different atmospheric metallicities. For the high atmospheric metallicity case, the gap in the $f_{env}$ distribution is located between 0.5 and 0.7, which is consistent with assumptions of the core-accretion model in which runaway accretion starts when $M_{env} \approx M_{core}$ ($f_{env}$ is $\sim 0.5$). We also find a bimodal distribution of the hydrogen and helium mass fraction ($f_{H/He}$) with a gap at $f_{H/He} = 0.3$. The location of this gap is independent of the assumed atmospheric metallicity. 
 Lastly, we compared the distributions of our sub-Saturns in the Neptunian savanna to a population of sub-Saturns in the Neptune desert and ridge. We find that the observed $f_{env}$ distribution of savanna and ridge sub-Saturns is consistent with the planets coming from the same underlying population.}


   \maketitle

\section{Introduction}
The prevailing theory to explain the formation of gas-rich planets is the core-accretion model  \citep{Mizuno1978,Mizuno1980,Stevenson1982,Bodenheimer1968,Lissauer1993,Pollack1996}. In this paradigm, solid cores with several times the Earth's mass ($\sim 10 -20~\text{M}_\oplus$) accrete gas from the surrounding protoplanetary disk. Initially, the accretion rate is set by the atmosphere's cooling time \citep{Lee2015,Ginzburg2016}. After the mass of the gaseous envelope reaches the mass of the core, the self-gravity of the atmosphere is no longer negligible and the planet enters a stage of runaway accretion, when it is able to accrete gas at a much faster rate \citep{Perri1974,Mizuno1980,Stevenson1982,Pollack1996,Lee2014}.

Under this assumption of core accretion, gaseous planets can be divided into two groups. The first comprises planets that never reached an envelope mass fraction ($f_{env} = M_{env}/(M_{env} + M_{core})$) of 50~\% and thus do not start runaway accretion. These planets would likely have an internal structure similar to that of Neptune and Uranus \citep[$f_{env} \sim 10 -25~\%$,][]{Miguel2023}. The second is gas giants that started runaway accretion and acquired a massive envelope ($f_{env}$ of $\sim 80 - 100~\%$), with internal structures similar to those of Jupiter and Saturn \citep{Miguel2023}.
Planets with $f_{env}$ of $\sim 50 - 70$~\% should be rare as they would have started the process of runaway accretion but then somehow quickly stopped accreting, leading to only a moderate gain in envelope mass \citep{Ida2004,Mordasini2009}. While these planets could have started runaway accretion just shortly before the protoplanetary disk dispersed, the required coincidence of these two rapid processes makes it unlikely.

As the population of known exoplanets grows, we can look for imprints of the core-accretion model on the population of gaseous planets. Intermediate-sized exoplanets ($4~\text{R}_\oplus< R_p < 8~\text{R}_\oplus$), often called super-Neptunes or sub-Saturns \citep[hereinafter sub-Saturns;][]{Hartman2009,Petigura2017}, are key targets to study, as we expect the transition to runaway gas accretion to happen in this regime. Short-period sub-Saturns are well studied, with a known paucity of sub-Saturns on very short orbits, called the Neptune desert \citep{lecavelier2007,szabo2011,nesvorn2013,ojeda2014,lundkvist2016,Mazeh2016}. 
The responsible mechanism for this desert is more likely connected to the long-term evolution of these planets due to the proximity to their stellar hosts, rather than their formation. The most favored explanations include a combination of atmospheric escape \citep{kurokawa2014,jackson2017,owen2019,koskinen2022} and migration mechanisms \citep{koenig2016,owenlai2018,viss2022}.

Recently, \cite{Bourrier2023} reported an underdensity of sub-Saturns even with longer orbital periods, which they named the Neptunian savanna. The Neptunian savanna is separated from the Neptune desert by an overdensity of sub-Saturns with periods between 3.2 and 5.7 days, coined the Neptunian ridge \citep{castro2024a}, which coincides with the hot-Jupiter pileup \citep{udry2003}. Given that planets in the savanna are influenced less strongly by their host stars, it could be a remnant of formation through core accretion and runaway accretion. However, as a large portion of these planets are expected to form beyond the ice line and then migrate inward, it is unclear how well the observed population represents their primordial distribution. If runaway accretion is indeed responsible for the deficit of sub-Saturns with longer periods, this should be reflected in the distribution of $f_{env}$.

The envelope mass fraction can be estimated from the observed planetary masses and radii through models of their interior structure. A two-component model composed of an Earth-composition core and a H/He envelope \citep{Lopez2014} has been used to derive individual envelope mass fractions of sub-Saturns \citep{Petigura2016,Petigura2017,Thomas2025}. These models found $f_{env}$ to be $\geq 50~\%$ for some of these planets, which should be unlikely according to core-accretion and runaway gas accretion theories \citep{Petigura2018a,Petigura2020,Mistry2023}. \cite{Millholland2020} show that for close-in sub-Saturns, tidal heating can inflate the radii of these planets, causing these simple models to overestimate $f_{env}$ by up to 60~\%. After considering tidal inflation, most close-in sub-Saturns in their sample had $f_{env}$ similar to that of Neptune. Although the exact reduction in envelope fraction depends on a couple of unknowns, such as the tidal quality factor and the obliquity \citep{Millholland2020}.

Another limitation of previous studies that used internal structure models to estimate $f_{env}$ is the assumption of a pure H/He envelope, which is inconsistent with the Solar System gas planets and their varying levels of envelope enrichment \citep{Irwin2014,Guillot2015}. Additionally, some sub-Saturns have been shown to have high-metallicity atmospheres \citep{MacDonald2019}, and both interior structure modeling \citep{Thorngren2016} and planet formation models \citep{Cridland2020b,Danti2023} suggest that planets in this mass range should be metal-rich. During envelope accretion, solid objects in the form of pebbles and/or planetesimals are accreted along with the gas. If the gaseous envelope is sufficiently dense to evaporate these bodies before they reach the core, the atmosphere will be polluted with metals \citep{Pollack1986,Alibert2017,Lozovsky2017,Brouwers2018}. Envelope enrichment significantly influences the formation of giant planets by reducing the critical core mass for runaway accretion and thus shortening the timescales for giant planet formation \citep{Ikoma2000,Hori2011,Venturini2016,Valletta2020}. On the other hand, \cite{Alibert2018} suggest that an intermediate stage of efficient heavy element accretion will delay the onset of runaway accretion until $M_p$ reaches $\sim 100~\text{M}_\oplus$. This could explain the observed differences in bulk metallicity between Saturn and Jupiter since Saturn would not have started runaway accretion in this case \citep{Venturini2020,Helled2023}.

In this work, we analyzed the population of warm sub-Saturns to connect their observed properties to planet formation theory. We used the GAS gianT modeL for Interiors \citep[GASTLI;][]{acuna2024,acuna2021} interior structure modeling code\footnote{https://github.com/lorenaacuna/GASTLI} to derive envelope mass fractions for our sample. GASTLI expands the simple two-component model by taking the atmospheric metallicity into account for a more accurate calculation of the planet's interior structure. Using GASTLI, we calculated a grid of interior-structure models on which we performed interior structure retrievals on our sample of sub-Saturns to analyze the distribution of $f_{env}$. In Sect. \ref{sampleselection} we present the selection of our sub-Saturn sample and how we recomputed the ages for all the stars in this sample. Section \ref{gastlisection} describes the computation of the interior structure model grid on which the retrievals were performed. The results are presented in Sect. \ref{results}. In Sect. \ref{discussion} we compare our results to the planet formation model of \cite{APC2020} and discuss the observed distribution of $f_{env}$ in the context of core accretion.


\section{Sample selection} \label{sampleselection}
\subsection{Planet selection}
 The planets in our sample were selected according to the following criteria to enable modeling with GASTLI and minimize the effects of interaction between the star and planet. We used the NASA Exoplanet Archive \citep{Christiansen2025} composite table of confirmed planets.
\begin{enumerate}
    \item The planets need to have measured radii and masses for the retrieval. We limited our sample to planets in the radius range between 4 and 10~R$_\oplus$. Typically, sub-Saturns are defined between 4 and 8~R$_\oplus$. We extended this radius range to include a few planets in the Jovian planet population that are certain to have undergone runaway accretion to compare with the sub-Saturns. The lower limit on the mass of our targets ($M_p > 15$\,M$_\oplus$) is driven by GASTLI's default atmospheric grid \citep[see][for details]{acuna2024}. We did not impose an upper mass limit; however, all the planets in the final sample have masses of $\leq 1~\text{M}_J$. The planetary radius is much more sensitive to the envelope mass fraction than to the mass. As such, we required a radius measurement precision of $\sigma_R< 20\%$. In order not to bias the sample too much, and in an effort to maintain meaningful number statistics, we did not impose strict requirements on the mass precision and included all planets with a true mass measurement (i.e., no upper limits). 
        
    \item Orbital periods longer than 15 days. \cite{Millholland2020} show that the tidal forces that close-in planets are subjected to can cause radius inflation, which in turn leads to an overestimation of $f_{env}$. This effect is strongest in planets with $P < 10$~days and additionally depends on the eccentricity and obliquity of the planet. To limit the influence of tidal inflation on our derivation of $f_{env}$, we imposed an orbital period limit on the sample. The only planet in our sample with an eccentricity larger than 0.5 is TOI-2134~c ($e=0.67^{+0.05}_{-0.06}$). However, given the large period of 95 days, we do not expect tidal inflation to be strong enough to significantly affect the derived $f_{env}$.

    \item Planet equilibrium temperature <1000\,K. The atmospheric grid we used for GASTLI is only valid up to equilibrium temperatures of 1000~K. The equilibrium temperatures were calculated following Eq. 23 in \cite{acuna2024}, assuming a global average ($f_{av} = 4$).
\end{enumerate}

\subsection{Age determination} \label{agedet}
A key parameter to perform interior structure retrievals from observed radii and masses is the age of the planet. Planets contract over their lifetime as they cool after their formation. The change in radius is strongest in the first 1 - 2 Gyr, which makes it difficult to infer the interior structure from observed properties for young planets. To derive accurate envelope mass fractions for young planets, the age should ideally be known with a precision of 10~\% \citep{muller2023}. For planets older than 2 Gyr, the precision requirement can be relaxed. We calculated the ages for all planets in the sample using the Python code \texttt{stardate} \citep{stardate}, which combines isochrone fitting with gyrochronology to infer precise stellar ages. We used the stellar effective temperature ($T_{\mathrm{eff}}$), the metallicity (Fe/H), and broadband photometry from \textit{Gaia} (G, G$_{RP}$, and G$_{BP}$) and 2MASS (J, H, and K) as input for the isochrone fitting. For the gyrochronology, we estimated the rotation period (P$_{rot}$) of the star from spectroscopic $v~\sin(i)$ measurements and the stellar radii. Since the measured $v~\sin(i)$ is only a lower limit for the rotational velocity of the star, the calculated P$_{rot}$ only gives an upper limit. For some planets, P$_{rot}$ could be determined from brightness modulations in their light curves. We ensured that our calculated P$_{rot}$ matches the photometric values in these cases. The calculated ages for all stars are listed in Table \ref{tab:stars}. We also list the literature ages of the stars where available.

To ensure that we could derive accurate $f_{env}$, we limited the sub-Saturn sample to planets that are older than 1 Gyr within 1$\sigma$. The average uncertainty of the ages derived in this work (1.1~Gyrs) is smaller than that of the literature ages (1.9~Gyrs). From the 28 stars in our sample that had literature ages, only 6 did not agree within their $1\sigma$ uncertainties, and only K2-24 does not agree within $2\sigma$. Given the large error on the rotational period, it is possible that \texttt{stardate} might overestimate the age of the star compared to a calculation method that does not include gyrochronology. However, since both the literature and the \texttt{stardate} age are well above 1~Gyr, this discrepancy should not affect the derived $f_{env}$. Of the stars that made it into the sample, only Kepler-9 would have been excluded based on the literature age. On the other Hand, there are six stars that were excluded based on their re-derived ages, even though the literature ages were within our requirements. This means that there is likely no contamination from young stars in our sample, which could skew the results. However, it is possible that some of the excluded planets could in fact be older and therefore should have been included in the sample.

\begin{figure}
    \centering
    \includegraphics[width=0.95\columnwidth]{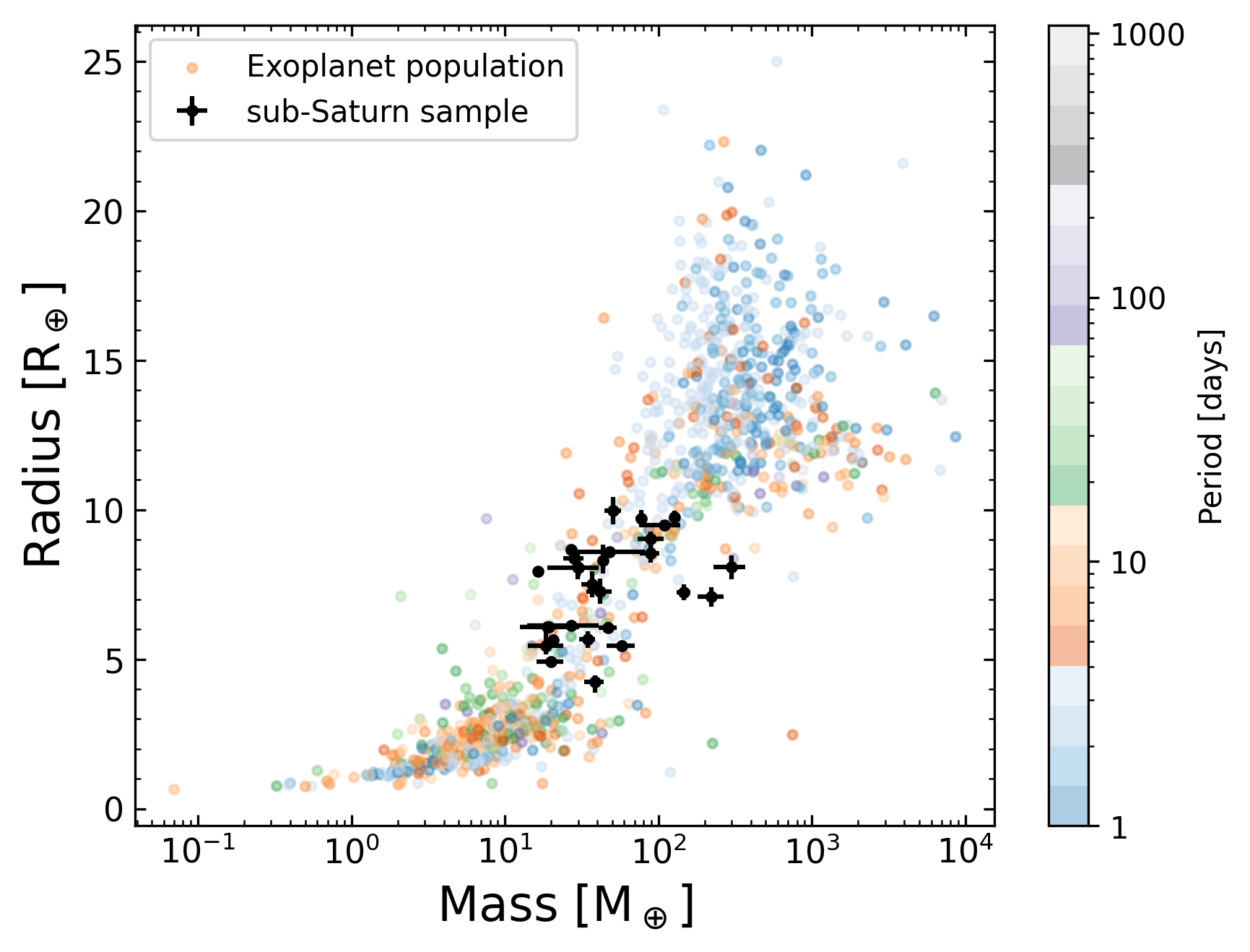} 
    \caption{Mass-radius diagram of the sub-Saturns in our sample. The colored points show the exoplanet population with radii and masses measured to the precision of our selection criteria. Their colors indicate the orbital periods of the planets. The sub-Saturns that were analyzed in this work are overplotted in black.}
    \label{fig:mr}
\end{figure}

In total, we were left with 28 sub-Saturns in our sample. They are all listed in Table \ref{tab:planets}. The majority of planets have orbital periods of $<50$~days due to the increasingly low transit probability at larger separations. Figure \ref{fig:mr} shows a mass-radius diagram of the planets in our sample in the context of the larger exoplanet population. The radii of the planets in our sample are evenly distributed over the whole range of sub-Saturn radii with a small gap between 6 and 7~R$_\oplus$. There are three high-density planets in our sample that are isolated compared to the rest of the population.

\section{Interior structure retrieval} \label{gastlisection}
\subsection{Interior-atmosphere model grid}
Using GASTLI, we first calculated a grid of interior structure models on which to perform retrievals to derive envelope mass fractions for our sub-Saturn sample. GASTLI solves the differential equations of planetary interior structure on a one-dimensional grid, assuming conservation of mass, hydrostatic equilibrium, convection, and thermal evolution due to secular cooling. It uses state-of-the-art equations of state for H/He, water, and rock \citep[see the references in][]{acuna2024} to compute density and entropy. GASTLI assumes two compositional layers: a metal-rich core and an envelope. The core composition is fixed to a 1:1 mass ratio of water and rock \citep[see also][]{Thorngren2016}, while the envelope consists of H/He with water as a proxy for metals. The mass fraction of metals in the envelope, $Z_{env}$, is defined as a user input parameter.

The interior structure model is coupled to a grid of atmospheric models via an iterative algorithm \citep{acuna2021}. This approach allows the temperature at the interior-atmosphere boundary to be calculated self-consistently. We also used the atmospheric pressure-temperature profile to integrate the equation of hydrostatic equilibrium and compute the atmospheric thickness, which contributes to the total planetary radius. For this calculation, we assumed a transit pressure of 20 mbar, which corresponds to the pressure level probed by transit photometry \citep{Grimm18}. Further details of the implementation are given in \cite{acuna2024,gastli_software}. GASTLI uses a default atmospheric grid generated with petitCODE. PetitCODE solves the radiative transfer equation and applies the Schwarzschild criterion to obtain self-consistent atmospheric pressure-temperature profiles and Bond albedos. More details on petitCODE, including its atmospheric species and the references of their opacity tables, can be found in \cite{molliere15,molliere17}.

The interior-atmosphere models in GASTLI take the following inputs: planetary mass ($M_p$), internal temperature ($T_{int}$), equilibrium temperature ($T_{eq}$), core mass fraction (CMF = $1- f_{env}$), and atmospheric metallicity ($\log(\mathrm{Fe}/\mathrm{H})$). The equilibrium temperature is calculated as a global average and assuming a null Bond albedo, consistent with the input of the petitCODE grid \citep[see Eq. 23 in][]{acuna2024}. The effective temperature of the atmosphere is given by $T_{eff}^{4} = T_{int}^{4} + (1-A_{B}) \times T_{eq}^{4}$, where $T_{int}$ is an input, and the Bond albedo ($A_{B}$) is computed self-consistently by petitCODE. The planet's internal heat is $E_{int} = \sigma_{SB} \times T_{int}^{4}$, where $\sigma_{SB}$ is the Stefan-Boltzmann constant. Given these five input parameters -- $M_p$, $T_{int}$, $T_{eq}$, CMF and $\log(\mathrm{Fe}/\mathrm{H})$ -- the model output parameters are the total radius ($R_{tot}$), the age, the total metal mass fraction ($Z_{planet}$ = $\text{CMF} + f_{env} \times Z_{env}$), and the atmospheric metal mass fraction ($Z_{env}$). $Z_{env}$ is equivalent to  $\log(\mathrm{Fe}/\mathrm{H})$ but expressed as a mass fraction instead of solar units.

The input masses in our grid range between 0.05~M$_J$ and 1.05~M$_J$. For the lower masses, the grid has a finer spacing of 0.01~M$_J$ as the absolute errors in the planet masses are smaller. Between 0.2~M$_J$ and 0.5~M$_J$ grid points are 0.03~M$_J$ apart. From 0.5~M$_J$ until 1.05~M$_J$ we used a spacing of 0.05~M$_J$. The equilibrium temperature grid spans values from 300 to 1000~K in steps of 50~K. For the atmospheric metallicity, we used values of -2.0, 0.0, 1.0, and 1.7 dex. We calculated the internal structure at internal temperatures of 50~K, 100~K, 200~K, 300~K, and 400~K.

The lower limits of planetary mass and atmospheric metallicity are set by GASTLI's atmospheric grid. Since our sample consists only of mature planets, the internal temperature range covered by the grid is sufficient to encompass their ages. The equilibrium temperatures in our sample range from 295\,K to 999\,K. The upper limit on atmospheric metallicity, set at $50~\times$~solar, is motivated by two factors. First, this value corresponds to the highest envelope metallicity observed in Neptune-mass exoplanets \citep{wakeford18}, while Neptune has an atmospheric metallicity of $80~\times$~solar. Thus, more massive exoplanets in our sample are expected to have metallicities lower than 50-80 $\times$ solar. Second, GASTLI's default atmospheric grid extends the interior-atmosphere interface pressure ($P_{surf}$) to 1000\,bar for $\log(\mathrm{Fe}/\mathrm{H})$ $<$ 2. At higher metallicities, $P_{surf}$ must be reduced to $\sim$10\,bar. This pressure range ensures convergence of the petitCODE atmospheric models. Moreover, at these pressures, the atmosphere is convective due to the high opacity, such that the adiabatic (convective) temperature profile computed by GASTLI avoids spurious temperature jumps at the interior–atmosphere boundary.

\begin{figure}
    \centering
    \includegraphics[width=0.95\columnwidth]{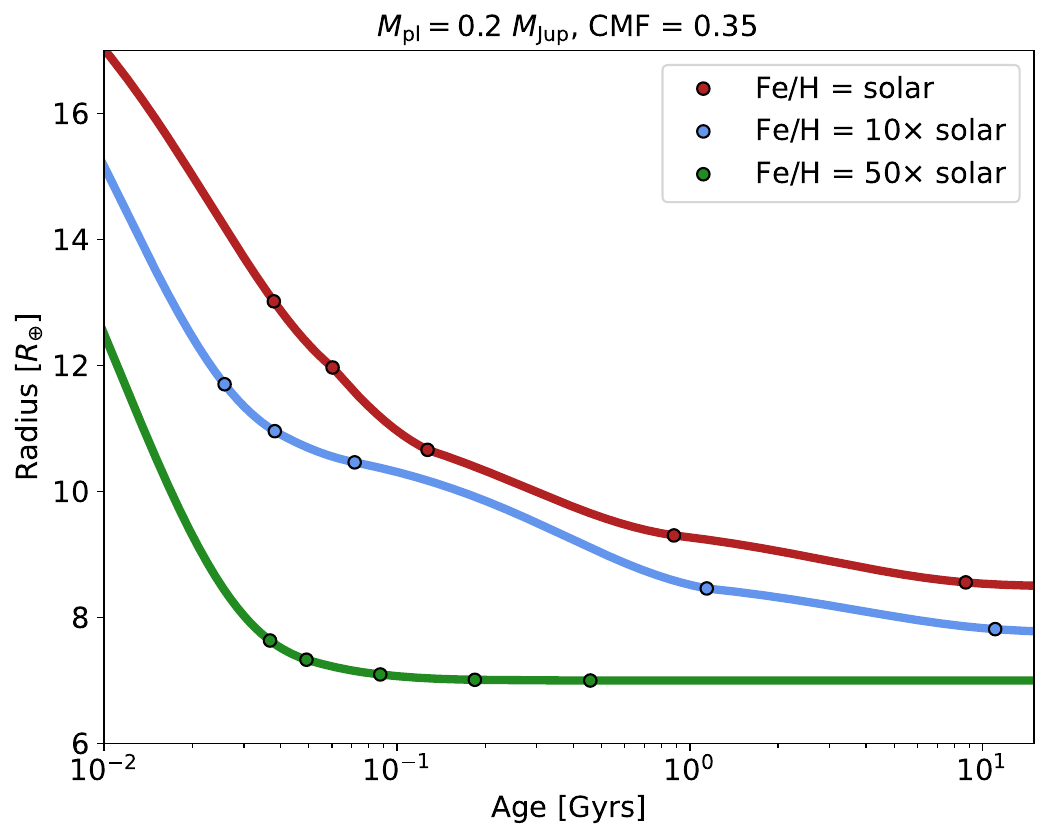} 
    \caption{Thermal evolution curves for a planet with a CMF of 0.35 and a mass of $0.2\, \mathrm{M}_{\mathrm{J}}$ with varying atmospheric metallicities. The input internal temperatures are converted to ages, and the planet radius at each step is calculated by GASTLI. A higher assumed atmospheric metallicity leads to a lower total radius. After $\sim 1$~Gyr, the change in radius due to the contraction of the planetary envelope is small even for low-metallicity atmospheres. }
    \label{fig:evo}
\end{figure}

This change in surface pressure can produce irregular entropy-age and radius-age curves, which require validation. Therefore, we adopted the $50~\times$~solar metallicity limit and validated the thermal evolution curves on a case-by-case basis. This involved identifying models in which the surface pressure changes. For those cases, we either adopted the values from the last stable point for the subsequent time steps or interpolated between two valid models. This approximation does not affect the forward models, as these occur at low internal temperatures, where the planetary radius remains constant with age (see Fig.~\ref{fig:evo}).

The validation of our GASTLI grid involves ensuring all models are in hydrostatic equilibrium. Some of the models in our grid with the lowest CMFs and planet masses are not in hydrostatic equilibrium due to the low surface gravity. For these models, GASTLI does not converge because they are unphysical. We consequently replaced these points with NaNs in our grid. This affected roughly 3~\% of the models. In total, our grid consists of 42,180 models. We have made the grid publicly available via Zenodo (\url{https://zenodo.org/records/17396266}).

\subsection{Retrieval setup}
Following the approach in \cite{acuna2024}, we applied a Markov chain Monte Carlo framework to retrieve posterior distributions for the interior structure parameters of the planets in our sample, using \texttt{emcee} \citep{mackey2013}. The log-likelihood was calculated using the radii, masses, and ages of our planets \citep[see Eqs. 6 and 14 in][]{Dorn2015,acuna2021}:
\begin{align}
    \log\mathcal{L}=-0.5 \left(\frac{(R_i-R_{obs})^2}{\sigma^2_{R_{obs}}}+\frac{(M_i-M_{obs})^2}{\sigma^2_{M_{obs}}}+\frac{(age_i-age_{obs})^2}{\sigma^2_{age_{obs}}}\right).
\end{align}
We found that there were only marginal differences in the results when fitting for the equilibrium temperature compared to running the retrieval with T$_{eq}$ fixed to the closest node in our grid. To reduce computational time, we therefore fixed T$_{eq}$ in the following analysis. The rest of the parameters are given uniform priors spanning the full range of our grid. We used 32 walkers with 200,000 steps while checking convergence using the autocorrelation time, $\tau$ (ensuring $\tau << N/50$).

\begin{figure}
    \centering
    \includegraphics[width=0.90\columnwidth]{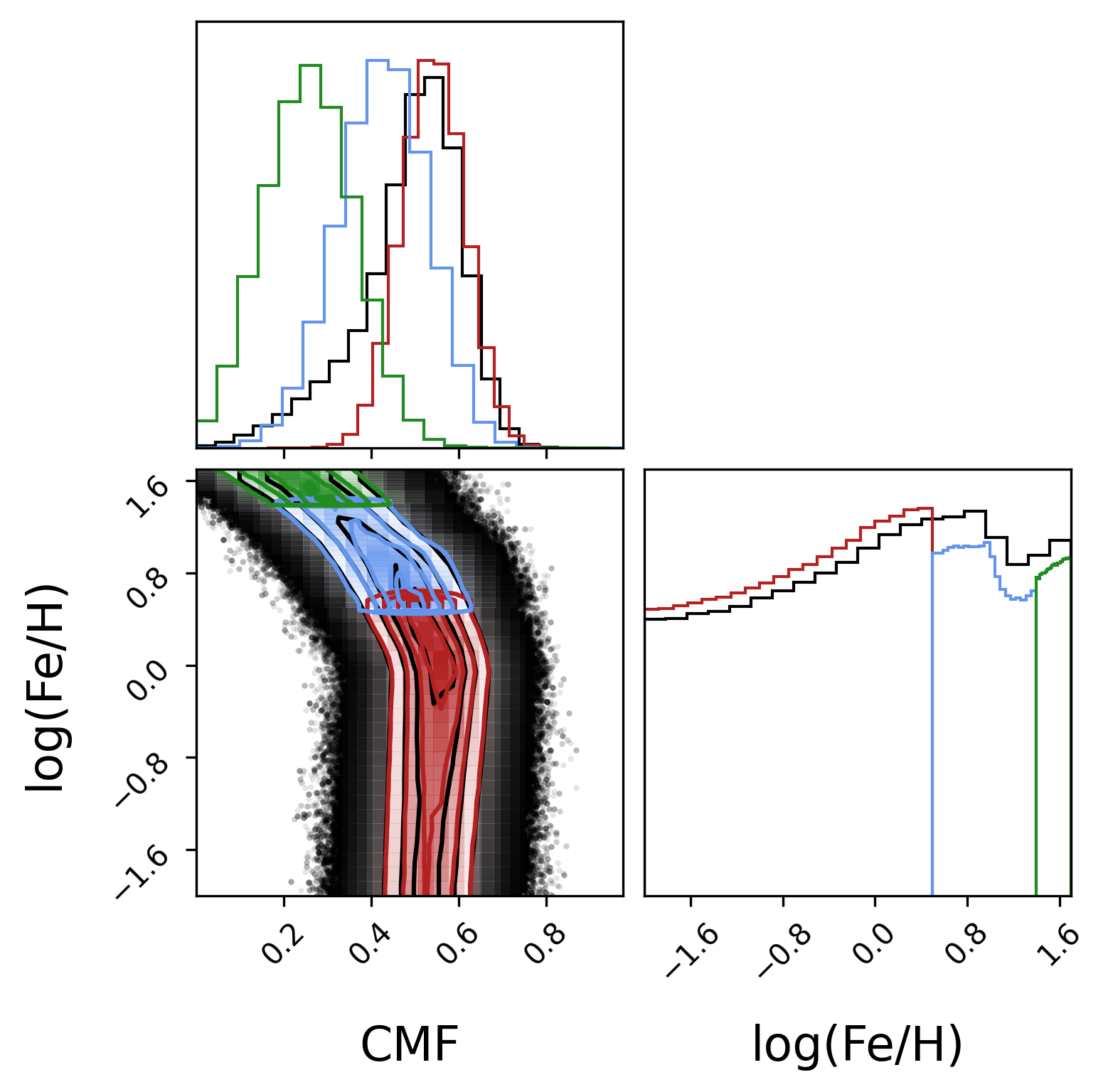} 
    \caption{Corner plots of the interior structure retrieval for Kepler-280\,b. Shown in black is the retrieval without any constraint on $\log(\mathrm{Fe}/\mathrm{H})$. While the CMF is uncorrelated for the majority of $\log(\mathrm{Fe}/\mathrm{H})$ values, for high atmospheric metallicities the inferred CMF decreases significantly. To illustrate this, the contours for low ($ -2.0 < \log(\mathrm{Fe}/\mathrm{H}) < 0.5$), medium ($ 0.5 < \log(\mathrm{Fe}/\mathrm{H}) < 1.4$), and high ($ 1.4 < \log(\mathrm{Fe}/\mathrm{H}) < 1.7$) atmospheric metallicity retrievals are overplotted in red, blue, and green, respectively.}
    \label{fig:deg}
\end{figure}

When running the retrievals on the full range of $\log(\mathrm{Fe}/\mathrm{H})$ values, we found that there is a strong degeneracy between the inferred envelope mass fraction and the atmospheric metallicity (see Fig.\,\ref{fig:deg}). For small atmospheric metallicities ($\log(\mathrm{Fe}/\mathrm{H}) < 0.5$), $f_{env}$ does not vary significantly with the atmospheric metallicity. However, for a high value of $\log(\mathrm{Fe}/\mathrm{H})$, the inferred $f_{env}$ can differ from the low-metallicity scenario by up to 0.3 - 0.4. This illustrates that prior constraints on the atmospheric metallicity are needed to derive accurate $f_{env}$ from interior structure models, assuming the age is known to a reasonable precision \citep[see the discussion in][]{acuna2024}.
However, none of the planets in our sample have $\log(\mathrm{Fe}/\mathrm{H})$ measurements.
To quantify how different atmospheric metallicities would change the inferred $f_{env}$ distribution in our sample, we ran retrievals for each planet for three metallicity cases: low ($ -2.0 < \log(\mathrm{Fe}/\mathrm{H}) < 0.5$), a medium ($ 0.5 < \log(\mathrm{Fe}/\mathrm{H}) < 1.4$), and high ($ 1.4 < \log(\mathrm{Fe}/\mathrm{H}) < 1.7$).

\section{Results} \label{results}
\begin{figure*}
    \centering
    \includegraphics[width=1.99\columnwidth]{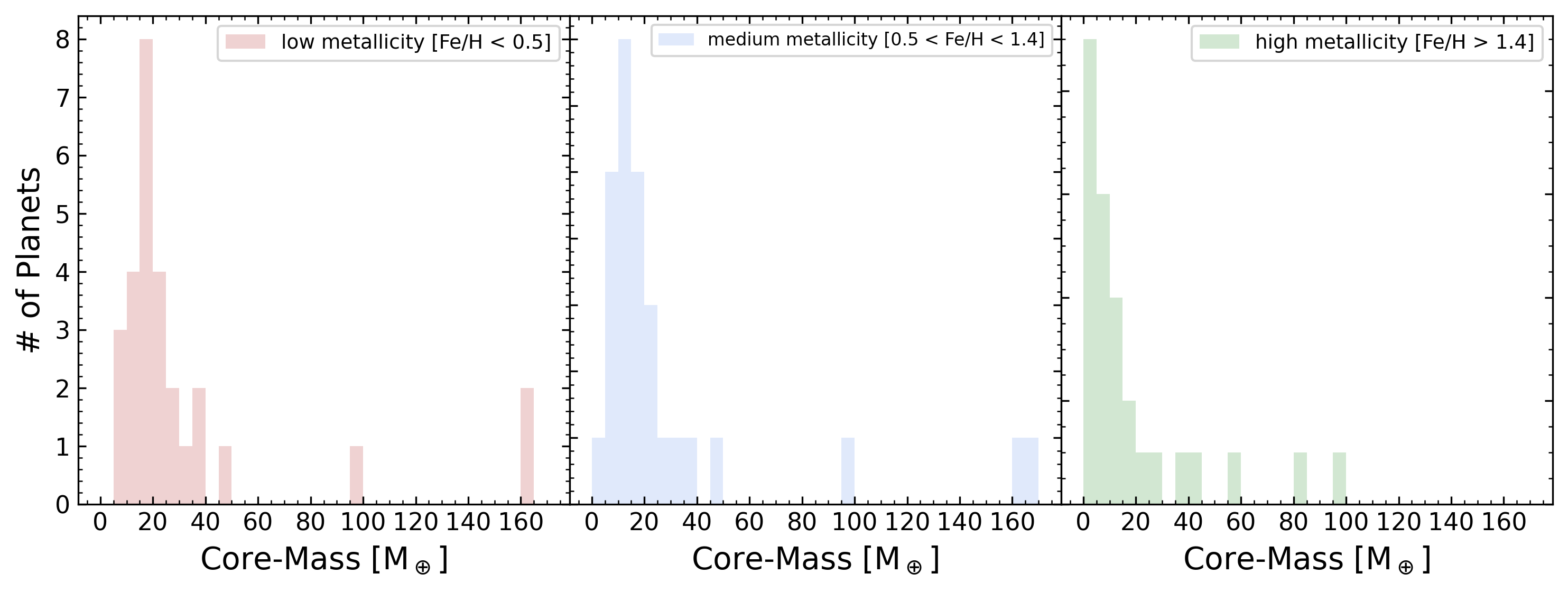} 
    \caption{Histograms showing the distribution of core masses for the planets in our sample. For low and medium atmospheric metallicities, the inferred core masses are mostly between $10$  and $30~\text{M}_\oplus$. The high $\log(\mathrm{Fe}/\mathrm{H})$ retrievals produce lower core masses. Three large outliers can be seen with core masses $>50~\text{M}_\oplus$.}
    \label{fig:cm}
\end{figure*}

The derived envelope mass fractions of all the planets in our sample for the three $\log(\mathrm{Fe}/\mathrm{H})$ cases are summarized in Table\,\ref{tab:planets}. As expected, $f_{env}$ is higher for larger $\log(\mathrm{Fe}/\mathrm{H})$ values of a given planet. The total core masses of the planets are shown in Fig.\,\ref{fig:cm}. The low and medium $\log(\mathrm{Fe}/\mathrm{H})$ cases show a similar distribution with most core masses between $10$ and $30~\text{M}_\oplus$. In the case of high $\log(\mathrm{Fe}/\mathrm{H}),$ the core mass distribution is shifted toward lower values ($>10\,\text{M}_\oplus$). 

In all three metallicity scenarios, there are three outliers of planets with unusually large core masses. These planets are TIC~139270665\,b, Kepler-111\,c, and Kepler-849~b. The masses of Kepler-111\,c and Kepler-849\,b reported in \cite{Dalba2024} have large uncertainties due to the long orbital period of the two planets and the relatively low number of available radial velocity observations. Additionally, as reported in \cite{Dalba2024}, there are hints of further planets in the systems, which would influence the accuracy of the reported masses. None of the other planets in our sample exhibited significant RV variations from potentially undetected companions. Given the unreliable masses, we excluded both planets from the following analysis.

For TIC~139270665~b, we noticed the stellar radius derived in \cite{Peluso2024} is considerably smaller than in the TESS Input Catalogue \citep{Paegert2022}. Considering that the derived parallax for TIC~139270665~b \citep[see Table\,1 in][]{Peluso2024} is significantly different from the \textit{Gaia} DR3 value, we found that the ExoFAST modeling from \cite{Peluso2024} conflated the stellar input parameters with those of another star (Dalba \& Peluso, private communication). Therefore, the true stellar radius is likely larger than previously assumed, which would also increase the true radius of TIC 139270665 b. Updated modeling of the stellar and planetary parameters found a radius of $9.8 \pm 0.5~\text{R}_\oplus$ (Dalba \& Peluso, private communication). With this new radius, we ran a new set of retrievals for the planet.

Without Kepler-111\,c and Kepler-849\,b, we have a total of 26 planets with derived $f_{env}$ for our analysis. The average uncertainty for $f_{env}$ in our sample is 0.05, 0.07, and 0.05 for the low, medium, and high $\log(\mathrm{Fe}/\mathrm{H})$ retrievals, respectively. We observed strong correlations between the percentage uncertainty of the radius and the error on the derived envelope mass fraction. We also find a correlation between the percentage uncertainty of the age and the associated $f_{env}$ uncertainty, albeit weaker than the correlation with the radius uncertainty. Although the observed radius does not change much after 1~Gyr, as shown in Fig. \ref{fig:evo}, there is still a small slope for some of the models. As such, a larger age uncertainty increases the uncertainty of $f_{env}$. The mass uncertainty, on the other hand, showed only a weak correlation.

\section{Discussion} \label{discussion}

\subsection{$f_{env}$ distribution of sub-Saturns}
We compared the results from our interior modeling to a population of synthetic planets from the core-accretion model of \cite{APC2020}. That model followed the semi-analytic formation model of \citep{Ida2004}, which grows the planetary core through successive collisions of planetesimals from their underlying (assumed) smoothly distributed disk. The growing embryos gravitationally interact with their natal disk and migrate toward their host star throughout their formation, following the "planet trap" migration model \citep[see][ for details]{Cridland2019a}. When the influx of planetesimals is sufficiently small, the growing embryos begin to accrete a gaseous envelope. In that model, this transition occurs near a core mass of $\sim 5 - 10$ M$_\oplus$; however, a few planets accreted cores of up to 50 M$_\oplus$ prior to the accretion of their gas\footnote{The precise core mass for initiating gas accretion depends on the choice of a threshold planetesimal accretion rate of one M$_\oplus$ per Myr, based on the results of \cite{Ikoma2000}.}. To conform to the selected sample of this work, we limited the population of synthetic planets to the same period restriction as described above. Unlike the observed population, the synthetic planets are considered "young" and their interior structure is not computed (more below). Since we did not infer their physical size, only their mass and their $f_{env}$ (derived directly from their accretion history) were compared.

For simplicity, their gas accretion rate is limited by a parameterized form of the Kelvin-Helmholtz timescale \citep[see][]{Ida2004}, with parameters that fit the numerical simulations of \cite{Ikoma2000}. The functional form of the Kelvin-Helmholtz timescale used in \cite{APC2020} has $\tau \propto M^{-2}$, so that when gas accretion first begins the accretion is slow ($\tau \sim 0.5$ Myr), but by the time the planetary embryo has grown to a Saturn-mass the accretion timescale has dropped to $\tau \sim 5$ kyr. While this parameterization captures the expected evolution of gas accretion, it does not explicitly handle the physical change (the envelope becoming self-gravitating) that is expected to cause the runaway accretion discussed here.

\begin{figure*}
    \centering
    \includegraphics[width=1.99\columnwidth]{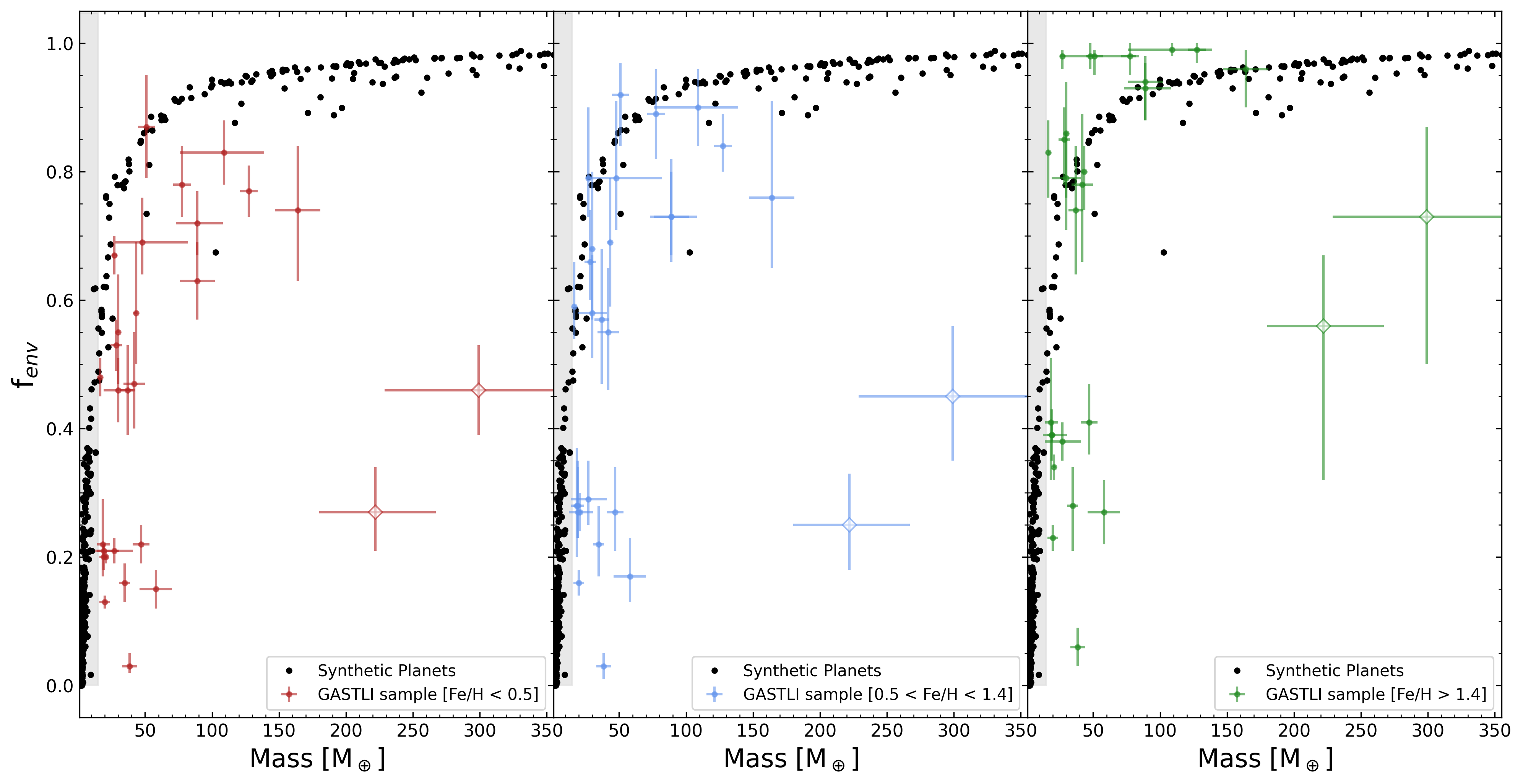} 
    \caption{Distribution of $f_{env}$ as a function of planet mass for the sub-Saturns in our sample. We derived three different $f_{env}$ for a low (left panel), medium (middle panel), and high (right panel) atmospheric metallicity case. The two planets with unreliable mass or radius measurements are indicated by the diamond markers. Black points are synthetic planets from the planetesimal accretion formation model of \cite{APC2020}. The gray shaded area indicates planets with masses below the mass limit of our sample. The synthetic planets represent young planets at a time just after their natal disk has photo-evaporated. While they will contract their envelope over time, this does not affect the total mass and envelope mass fraction.}
    \label{fig:synthcomp}
\end{figure*}

Figure \ref{fig:synthcomp} shows $f_{env}$ as a function of the planetary mass for the synthetic population and our sample. In the low $\log(\mathrm{Fe}/\mathrm{H})$ case, the envelope mass fractions in our sample are almost all significantly smaller than expected from the planet formation model for a given planet mass. The medium- and high-metallicity retrievals produce $f_{env}$ that are more similar to what we see in the formation model. In the intermediate mass range ($\sim 40 - 50~\text{M}_\oplus$) the high $\log(\mathrm{Fe}/\mathrm{H})$ retrievals produced the most similar $f_{env}$ results.
However, in the high $\log(\mathrm{Fe}/\mathrm{H})$ case, some of the derived $f_{env}$ for sub-Saturns with larger masses ($\sim 60 - 100~\text{M}_\oplus$) significantly exceed the values we expect from the formation model. Here, the medium $\log(\mathrm{Fe}/\mathrm{H})$ results provide a better fit to the synthetic planets.
This is consistent with observations of atmospheric metallicities from the solar system. For the gas-rich planets in the Solar System, we find a roughly linear decrease in atmospheric metallicity from Uranus and Neptune ($\log(\mathrm{Fe}/\mathrm{H}) \sim 80~\times$~solar) to Saturn ($\log(\mathrm{Fe}/\mathrm{H}) \sim 10~\times$~solar) and Jupiter ($\log(\mathrm{Fe}/\mathrm{H}) \sim 2-3~\times$~solar; \citealt{Kreidberg2014}). While there is some tentative evidence of such a trend in the exoplanet population, it has not been definitively confirmed \citep{Thorngren2016,Welbanks2019,Ohno2025}.

For the fraction of low-$f_{env}$ sub-Saturns in our sample, even the high $\log(\mathrm{Fe}/\mathrm{H})$ retrievals are not in agreement with the synthetic planet population. One potential explanation is that these planets have even higher atmospheric metallicities than what we modeled for here, which would shift their $f_{env}$ to higher values, closer to the synthetic planets. Since most of the low $f_{env}$ planets in our sample also have lower masses similar to those of Neptune and Uranus, it is possible that their $\log(\mathrm{Fe}/\mathrm{H})$ exceeds the maximum of $50 \times$ solar of our retrievals. Alternatively, given that the population of synthetic planets of \cite{APC2020} represents "young" planets that have just emerged from their natal protoplanetary disk, there could be mass loss mechanisms that reduce the envelope mass fraction of these planets over gigayear timescales. However, the envelope mass fraction of sub-Saturns is not expected to be significantly affected by photoevaporation, especially beyond 0.1~au \citep{Chen2016,Millholland2020}. To investigate the effect of photoevaporation on our sample, we followed the approach in \cite{Doyle2025}. We calculated the photoevaporation mass-loss timescale, $t_m$:
\begin{align}
    t_m = \frac{M_{env}}{\dot M_{env}},
\end{align}
where $M_{env}$ is the mass of the envelope and $\dot M_{env}$ is the mass-loss rate of the envelope calculated from \cite{Owen2017}:
\begin{align}
    \dot M_{env}= \eta\frac{\pi R_p^3 L_{HE}}{4 \pi a^2 G M_p}.
\end{align}
For the efficiency parameter ($\eta)$, we assumed 0.1, and the high-energy flux was calculated using the relation from \cite{Jackson2012} $L \approx10^{-3.5}L_\odot(M_\star/M_\odot)$. Planets with mass-loss timescales $t_m > 100~$Myr are assumed to be stable against photoevaporation. All of the planets in our sample have photo-evaporative timescales well above 1~Gyr, the lowest being Kepler-450~b with 26~Gyrs. We therefore conclude that photoevaporation should not have a significant impact on the sub-Saturns in our sample.

\subsection{A runaway accretion gap?}

\begin{figure*}
    \centering
    \includegraphics[width=1.99\columnwidth]{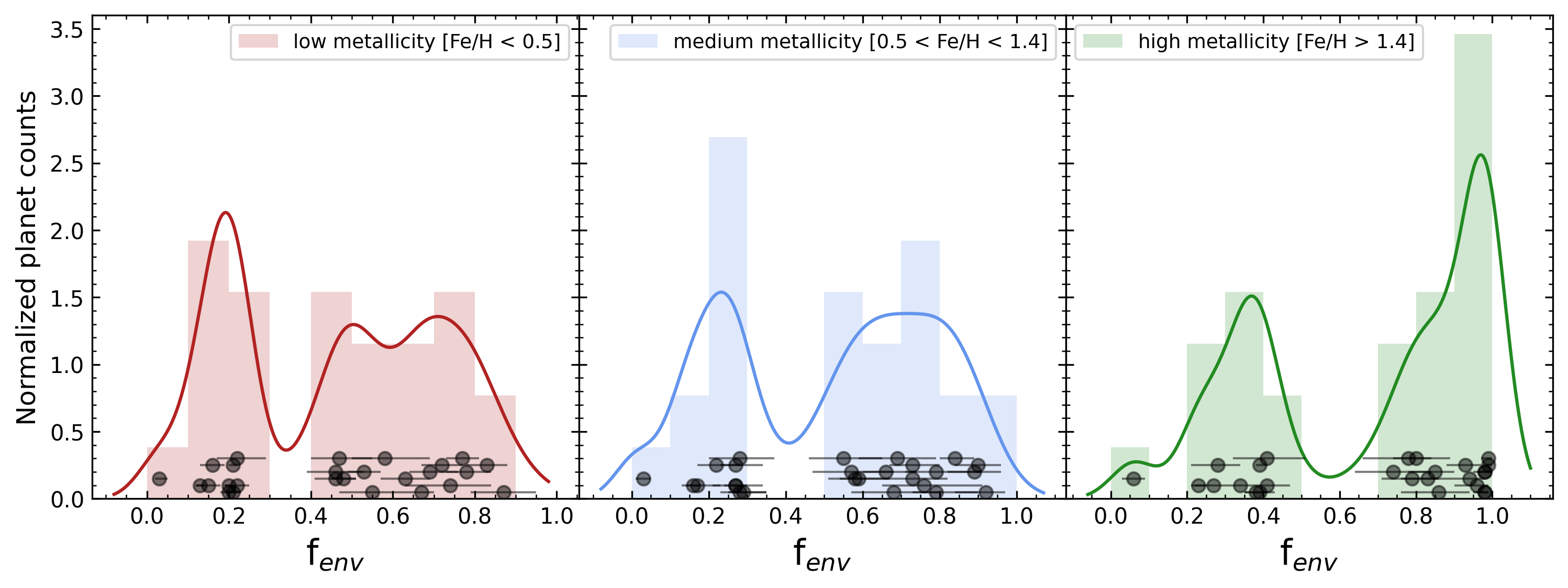} 
    \caption{Histograms showing the $f_{env}$ distribution for low (left panel), medium (middle panel), and high (right panel) atmospheric metallicities. The black points at the bottom are the inferred $f_{env}$ values of the planets in our sample (distributed in the y-axis direction for visibility). The solid lines show an estimation of the probability density of the $f_{env}$ distribution using kernel density estimation.}
    \label{fig:fenv}
\end{figure*}

Planets that start the process of runaway accretion will rapidly accumulate large amounts of gas. Assuming that runaway accretion starts when $M_{core} = M_{env}$, there should be a paucity of planets with $f_{env} \sim 50~\%$ as these planets would have to have started runaway accretion at nearly the same time as their disk evaporated and thereby only gaining a small amount of gas in their envelopes after reaching $f_{env} \sim 50\%$. To investigate the existence of such a "runaway accretion gap," we look at the distribution of $f_{env}$ values in our sample. Figure \ref{fig:fenv} shows histograms of $f_{env}$ for the three atmospheric metallicity cases. Together with the histograms, we plotted an estimate for the probability density function of the distribution using kernel density estimation. Kernel density estimation takes the measurement errors on the $f_{env}$ values into account and convolves them with a bandwidth that is used to control the smoothness of the estimate. For all three distributions, we used a bandwidth of 0.05.  

The three histograms have a bimodal distribution with a gap where no $f_{env}$ values are found. The location of the gap depends on the assumed metallicity of the atmosphere. For low and medium $\log(\mathrm{Fe}/\mathrm{H})$ atmospheres, the gap is between 0.3 and 0.5 $f_{env}$, with a slightly wider gap in the medium-metallicity case. The gap for the high $\log(\mathrm{Fe}/\mathrm{H})$ retrievals is between 0.5 and 0.7 $f_{env}$. This is what would be expected for the $f_{env}$ distribution after formation if runaway accretion starts at $M_{core} = M_{env}$. In this case, the intermediate-sized planets could be divided into super-Neptunes ($f_{env}  < 0.5 $) that did not start runaway accretion and sub-Saturns ($f_{env} > 0.7 $) that did.

\begin{figure*}
    \centering
    \includegraphics[width=1.99\columnwidth]{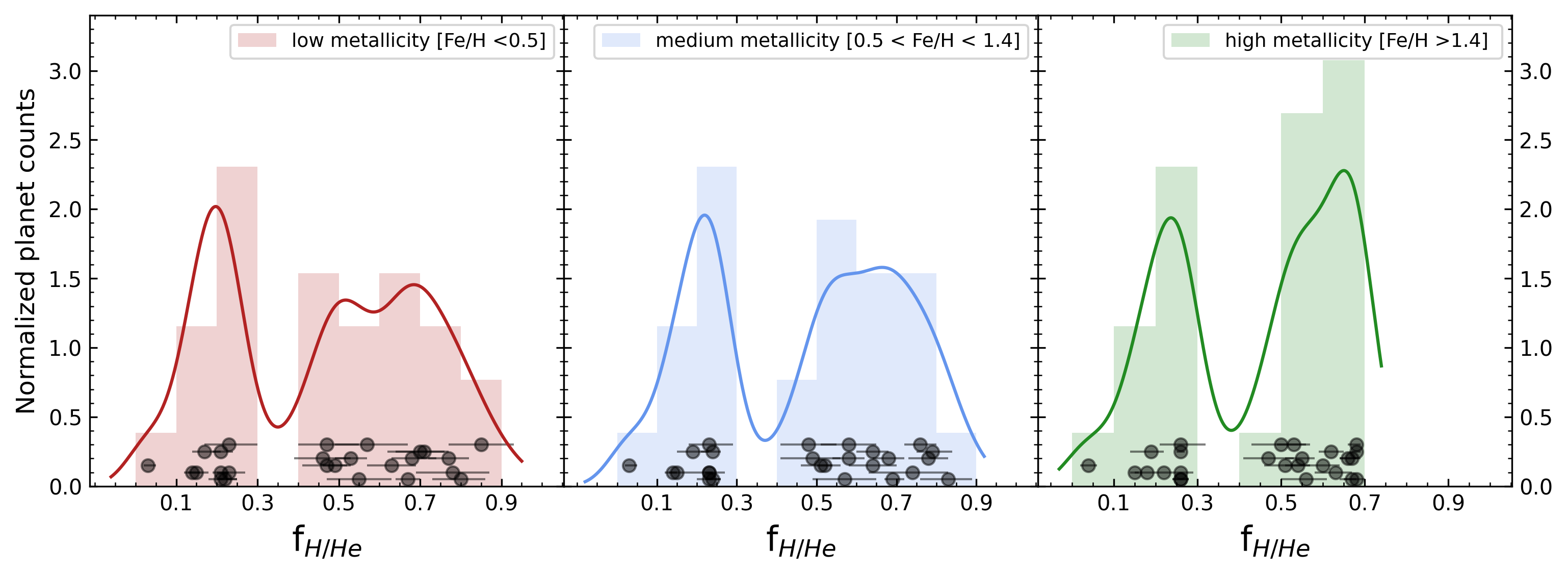} 
    \caption{Similar to Fig.\,\ref{fig:fenv} but showing the hydrogen and helium mass fraction ($f_{H/He}$) distribution for low (left panel), medium (middle panel), and high (right panel) atmospheric metallicities. The black points at the bottom are the inferred $f_{H/He}$ values of the planets in our sample (distributed in the y-axis direction for visibility).}
    \label{fig:fhhe}
\end{figure*}

\begin{figure*}
    \centering
    \includegraphics[width=0.95\columnwidth]{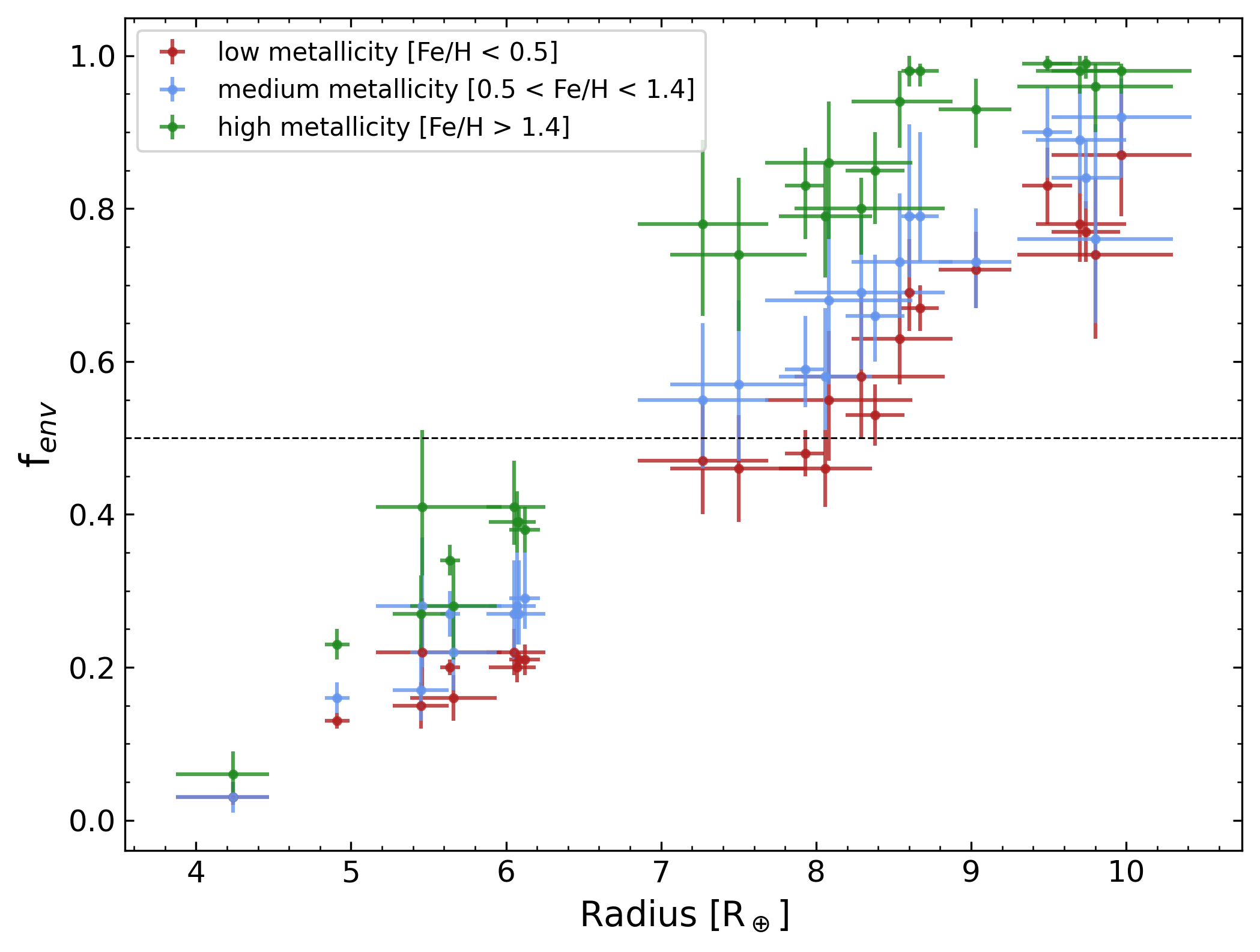} 
    \includegraphics[width=0.95\columnwidth]{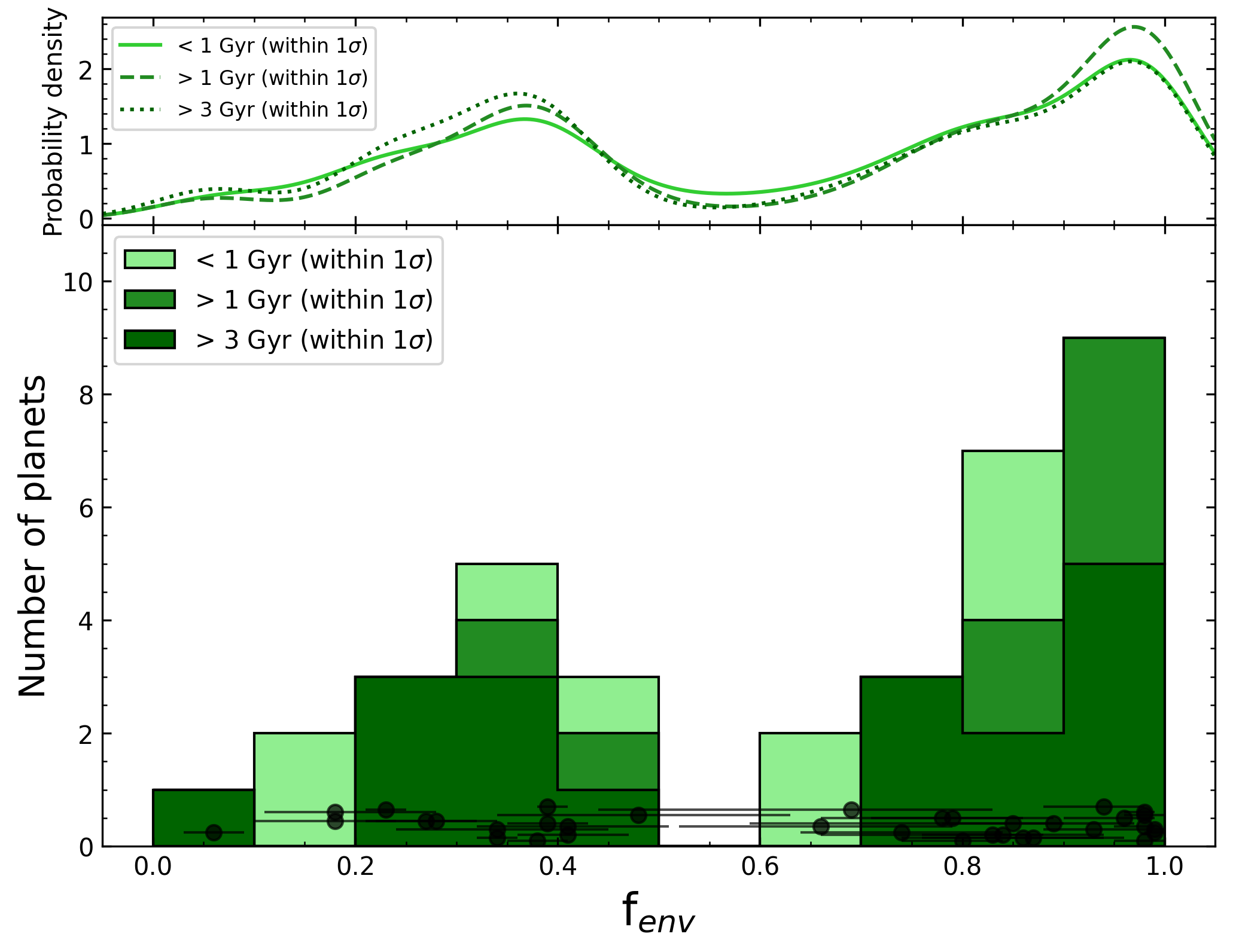} 
    \caption{Left: Derived envelope mass fraction as a function of radius for the planets in our sample. The dotted black line marks the 50~\% $f_{env}$ threshold. The gap in $f_{env}$ for the different retrievals coincides with a gap in radius between 6.2 and 7~R$_\oplus.$ Right: Distribution of $f_{env}$ as a histogram for different age thresholds. Planets are distributed in subsamples of planets older than 3~Gyrs within 1$\sigma$, planets older than 1~Gyr within 1$\sigma$, and all planets initially considered for the sample before the age cutoff. On top are kernel density estimator functions based on the subsample. The bimodal shape of the distribution is independent of the age threshold.}
    \label{fig:rad}
\end{figure*}

Another parameter that might be relevant for runaway accretion is the hydrogen and helium mass fraction ($f_{H/He}$). While $f_{env}$ encompasses the whole mass of the atmosphere, including metals, $f_{H/He}$ only considers the mass of hydrogen and helium in the atmosphere.
\cite{Venturini2016} computed self-consistent planet formation models including envelope enrichment and found that $f_{H/He} \sim 30~\%$ at the onset of runaway accretion for multiple different initial conditions. However, they did not find an obvious explanation for why that would be happening in their models. Looking at the distribution of $f_{H/He}$ in our sample, we find a gap between $f_{H/He}$ values of $0.3$ and $0.4$ (see Fig. \ref{fig:fhhe}), seemingly corroborating the results in \cite{Venturini2016}. While the exact shape of the distribution at higher $f_{H/He}$ changes, the position of the gap in $f_{H/He}$ at 0.3 is independent of the assumed atmospheric metallicity.
There is a clear gap in the distribution of both $f_{env}$ and $f_{H/He}$. For both distributions, the same planets fall below and above the gap. This bimodal distribution, including a "runaway accretion gap," is consistent with the model of core accretion, where runaway accretion is a rapid process that starts at $f_{env} \sim 0.5$. However, the gaps in the distributions of $f_{env}$ and $f_{H/He}$ could also stem from our limited sample size. As shown in Fig.~\ref{fig:rad}, the gap in $f_{env}$ coincides with a lack of planets in our sample with radii between 6.2~R$_\oplus$ and 7~R$_\oplus$. Given the strict selection criteria in our sample, we investigated whether this lack of planets exists in the wider exoplanet population. We still looked for mature, longer-period planets, as young or short-period planets could have radii in this range that are inflated, from either tides or internal heat, while still having low $f_{env}$ below the gap. In the NASA Exoplanet Archive composite table, there are a total of 80 sub-Saturns when imposing the same conditions on radius precision, orbital period, and age as we did for our sample, but without the requirement of a measured mass. Out of these 80 sub-Saturns, seven are listed in the composite data with a radius between 6.2~R$_\oplus$ and 7~R$_\oplus$. However, updated values for KOI-94~e and TOI-1386~b place them below 6.2~R$_\oplus$. The lack of planets in this radius regime may therefore be a real feature caused by the onset of runaway accretion. 

We also explored the impact of the age threshold on the distribution of $f_{env}$. Figure \ref{fig:rad} (right) shows a stacked histogram of the derived $f_{env}$ values assuming high $\log(\mathrm{Fe}/\mathrm{H})$ for different age bins. We compared how the distribution would change if we restricted the sample to even older planets (above 3~Gyrs within 1$\sigma$). For this comparison, we also included high $\log(\mathrm{Fe}/\mathrm{H})$ retrievals of the planets that were excluded after our age calculation because they were not older than 1~Gyr within 1$\sigma$. As shown in the histograms and the probability density functions on top of the plot, the distribution does not change significantly for the different ages. The distribution is bimodal in all three cases, although it is less pronounced if the young planets are included. In particular, there are two planets (Kepler-30~d and TOI-1823~b with $f_{env}$ between 0.6 and 0.7, narrowing the observed gap. However, given their young and uncertain age, the error bar on the derived $f_{env}$ is quite large. Additionally, as discussed briefly in Sect. \ref{agedet}, the literature ages for these two planets are older than the age derived in this work. Without the internal heat from the young age that was assumed for the retrieval, the envelope mass fraction would be larger to account for the observed radii moving the planets to the right of the histogram, thus widening the $f_{env}$ gap. 

\subsection{Differences between savanna sub-Saturns and desert and ridge sub-Saturns}
\begin{figure}
    \centering
    \includegraphics[width=0.90\columnwidth]{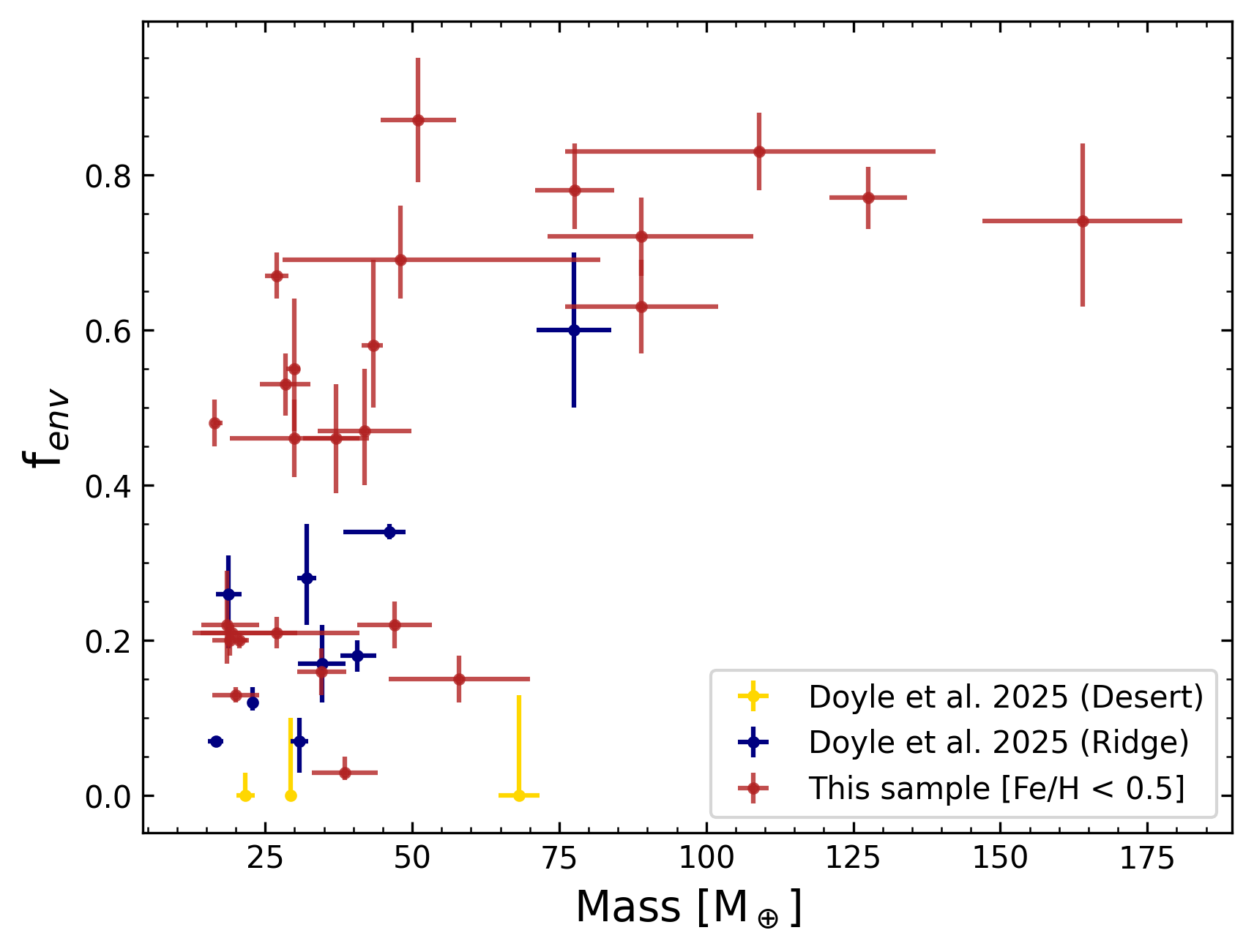} 
    \caption{Comparing the $f_{env}$ of our low $\log(\mathrm{Fe}/\mathrm{H})$ sample to the results from \cite{Doyle2025}. We only include planets that satisfy the sub-Saturn definition used in the work. The \cite{Doyle2025} sample is separated into planets within the Neptune desert ($P < 3.2$~d) and the ridge ($3.2~\text{d} < P < 6~\text{d}$).}
    \label{fig:doyle}
\end{figure}

\cite{Doyle2025} used GASTLI to analyze the envelope mass fractions of planets in the Neptune desert and ridge. In Fig.\,\ref{fig:doyle} we compare the $f_{env}$ distribution of desert and ridge sub-Saturns from \cite{Doyle2025} to our longer-period sample. In their work, they assumed solar metallicity in the $f_{env}$ retrievals, so to ensure comparability, we used our low $\log(\mathrm{Fe}/\mathrm{H})$ sample. Additionally, we only included planets from \cite{Doyle2025} that would be sub-Saturns in our sample (i.e., $4~\text{R}_\oplus<R_p <8~\text{R}_\oplus$). This removes the majority of planets in the Neptune desert, as these were mainly sub-Neptunes. The three sub-Saturns in the Neptune desert all have small envelope mass fractions. While this is expected for the two lower-mass planets, TOI-3071~b has an unusually low $f_{env}$ given its mass of $68 \pm 4~\text{M}_\oplus$ \citep{Hacker2024}. 

On the other hand, the distribution of $f_{env}$ for sub-Saturns in the Neptunian ridge is similar to the distribution of our Neptunian savanna sample. To assess the similarity between the $f_{env}$ distribution of the ridge sub-Saturns and our savanna sub-Saturns, we used a permutation test. Assuming that the two distributions come from the same underlying population, we partitioned the combined dataset randomly into two subsets corresponding to the original sizes of the two populations. We calculated the mean of the pairwise distance for 1000 permutations to compare with the observed distribution. Our test yields a p-value of 0.325, indicating no statistically significant difference between the two distributions. Our results indicate that there is no significant difference between the envelope mass fractions of sub-Saturns located in the Neptune ridge and those located in the Neptune savanna. Recent works have found potential differences in the density and host star metallicity between planets in the ridge and the savanna, pointing to different formation or evolution pathways of the two sub-Saturn populations \citep{castro2024b,Vissapragada2025}. While our results appear to be in contrast to their findings, the small sample size of the ridge sub-Saturns, particularly in the higher $f_{env}$ regime, makes it difficult to draw definitive conclusions.

\section{Summary and conclusions}
We derived the envelope mass fraction distribution of a sample of 28 warm sub-Saturns to compare with the outcomes from planet formation models. In particular, we investigated whether the onset of runaway accretion leaves an imprint on the population of sub-Saturns. We find that the exact distribution of $f_{env}$ in our sample depends strongly on the assumed atmospheric metallicity for the planets. The best agreement with predictions from the planetesimal accretion model of \cite{APC2020} is produced assuming medium ($0.5 < \log(\mathrm{Fe}/\mathrm{H}) < 1.4$) to high ($\log(\mathrm{Fe}/\mathrm{H}) > 1.4$) atmospheric metallicities.

Additionally, we find a bimodal distribution of $f_{env}$ with a gap that could be caused by the onset of runaway accretion. This bimodality is seen for all three assumed atmospheric metallicities. However, the location of this gap is dependent on the atmospheric metallicity and only matches theoretical predictions for high $\log(\mathrm{Fe}/\mathrm{H})$. This bimodality is also seen in the distribution of the hydrogen and helium mass fraction, $f_{H/He}$. The gap in this distribution is located at $30~\%$, corroborating the result from \cite{Venturini2016}, which is yet to be explained. Lastly, when comparing the distribution of $f_{env}$ for warm sub-Saturns in the Neptunian savanna to $f_{env}$ measurements from \cite{Doyle2025} of hot sub-Saturns in the Neptune desert and ridge, we find no statistically significant difference between the savanna and ridge population.

We have shown tentative evidence of a gap in the sub-Saturn population that may be caused by runaway accretion. However, we are limited by the size of our sample and the absence of measurements of atmospheric $\log(\mathrm{Fe}/\mathrm{H})$. Confirming more warm sub-Saturns  ($P > 15$~d) to expand the sample, particularly with radii between 6.2 and 7~R$_\oplus$, is key to further investigating the existence of a gap in the $f_{env}$ distribution. Additionally, measuring the atmospheric metallicities of these sub-Saturns will be crucial to breaking the degeneracy between $f_{env}$ and $\log(\mathrm{Fe}/\mathrm{H})$ and deriving accurate envelope mass fractions. This is especially important when interpreting the $f_{env}$ distribution in the context of planet formation models, such as linking the gap in $f_{env}$ to runaway accretion.

The $f_{env}$ is constant from subsolar $\log(\mathrm{Fe}/\mathrm{H})$ to $\sim 10 \times \text{solar}$ and increases rapidly for higher $\log(\mathrm{Fe}/\mathrm{H})$.\ As such, knowing in which regime a planet falls can help improve the derived $f_{env}$ even in the absence of extremely precise measurements. The results of an analysis like this can, in turn, inform planet formation models. 

\section{Data availability} \label{Sec:DataAvailability}
The model grids that are used to perform the interior structure retrievals are made publicly available via Zenodo (\url{https://zenodo.org/records/17396266}).
\begin{acknowledgements}
We thank the anonymous referee for the corrections and suggestions that helped us improve the presentation of the paper.
The authors thank Daniel Peluso, Paul Dalba, and Lauren Sgro for their input on the parameters of TIC~139270665\,b, and Jan-Niklas Pippert for helpful discussions surrounding the paper. LT acknowledges support from the Excellence Cluster ORIGINS funded by the Deutsche Forschungsgemeinschaft (DFG, German Research Foundation) under Germany's Excellence Strategy – EXC 2094 – 390783311. This research has made use of the NASA Exoplanet Archive, which is operated by the California Institute of Technology, under contract with the National Aeronautics and Space Administration under the Exoplanet Exploration Program.
\end{acknowledgements}

\bibliographystyle{aa}
\bibliography{bib} 
\begin{appendix}
\section{\texttt{stardate} results}

\begin{minipage}{\textwidth}
    \centering
    \resizebox{\textwidth}{!}{
    \begin{threeparttable}[b]
    \caption{\label{tab:stars} Input properties and ages from the \texttt{stardate} modeling of host stars. }
    \begin{tabular}{llllllll|c}
    \hline
    \hline
      Star   & $T_{\mathrm{eff}}$ [K] & Fe/H &$\text{R}_*~[\text{R}_\odot]$& Lit. Age [Gyr]&$v\sin(i)$  &$\text{P}_{rot}$ [days]& References  &Age~[Gyr] \\
      \hline

EPIC 211945201 & $6025 \pm 100$ & $0.1 \pm 0.1$ & $1.38^{+0.017}_{-0.018}$ &$3.99^{+0.85}_{-0.7}$& $2.93\pm 0.58$ & $24.0\pm 4.7$ & 1,2 & $4.4^{+0.7}_{-1.0}$  \\
K2-24 & $5726 \pm 65$ & $0.41 \pm 0.05$ & $1.185 \pm 0.011$ & $4.9 \pm 1.7$&$1.503\pm 1.0$ & $40.1\pm 27.0$ & 3,4 & $13.4^{+0.9}_{-0.4}$  \\
K2-280 & $5500 \pm 100$ & $0.33 \pm 0.08$ & $1.28 \pm 0.07$ & $8.96 \pm 1.7$&$2.6\pm 0.61$ & $25.1\pm 6.0$ & 5,2 & $10.2^{+7.0}_{-1.3}$  \\
KOI-1783 & $5922 \pm 60$ & $0.11 \pm 0.04$ & $1.14^{+0.04}_{-0.03}$ & -&$2.8\pm 1.0$ & $20.8\pm 7.4$ & 6,4 & $3.4 \pm 0.07$  \\
KOI-94 & $6184^{+110}_{-160}$ & $0.11 \pm 0.07$ & $1.43^{+0.09}_{-0.10}$ & - &$7.1\pm 0.77$ & $10.3\pm 1.3$ & 7,2 & $2.5^{+0.4}_{-0.7}$  \\
Kepler-103 & $6047^{+50}_{-70}$ & $0.15 \pm 0.04$ & $1.492^{+0.024}_{-0.022}$ & - &$3.34\pm 0.55$ & $22.7\pm 3.8$ & 8,2 & $5.2^{+0.5}_{-1.0}$  \\
Kepler-111 & $5914 \pm 90$ & $0.237 \pm 0.06$ & $1.145 \pm 0.04$ & $3.4^{+3.1}_{-2.2}$&$2.8\pm 1.0$ & $20.8\pm 7.5$ & 9,10 & $2.6^{+0.9}_{-0.8}$  \\
Kepler-117 & $6150 \pm 110$ & $-0.04 \pm 0.1$ & $1.61 \pm 0.05$ & $5.3 \pm 1.4$&$7.4\pm 1.0$ & $10.8\pm 1.5$ & 11,10 & $3.58^{+0.4}_{-0.27}$  \\
Kepler-1513 & $5491 \pm 100$ & $0.17 \pm 0.06$ & $0.95^{+0.08}_{-0.06}$ & $7.0^{+4.0}_{-4.2}$& $1.7\pm 0.5$ & $28.4\pm 8.7$ & 12,13 & $4.1 \pm 1.9$  \\
Kepler-450 & $6152 \pm 100$ & $0.13 \pm 0.02$ & $1.64^{+0.08}_{-0.11}$ & - &$13.4\pm 1.0$ & $6.2\pm 0.6$ & 7,4 & $3.67^{+0.11}_{-0.17}$  \\
Kepler-849 & $5950^{+140}_{-130}$ & $0.145 \pm 0.06$ & $1.82^{+0.07}_{-0.08}$ & $4.7^{+1.8}_{-1.6}$&$4.6\pm 0.52$ & $20.1\pm 2.4$ & 9,2 & $3.90^{+0.3}_{-0.23}$  \\
Kepler-9 & $5774 \pm 60$ & $0.05 \pm 0.07$ & $0.96 \pm 0.02$ & $2.0^{+2.0}_{-1.3}$&$2.78\pm 0.6$ & $17.5\pm 3.8$ & 14,2 & $2.5^{+0.9}_{-1.1}$  \\
TIC 139270665 & $6033 \pm 90$ & $0.19 \pm 0.06$ & $1.34 \pm 0.05$ & $3.7^{+2.1}_{-1.7}$&$4.2\pm 1.0$ & $16\pm 6$ & 15 & $3.8 \pm 0.5$  \\
TOI-1386 & $5793 \pm 80$ & $0.16 \pm 0.06$ & $1.02 \pm 0.03$ & $3.3^{+3.5}_{-2.2}$&$4.0\pm 2.0$ & $12.9\pm 6.5$ & 16 & $4.2^{+1.5}_{-1.2}$  \\
TOI-1439  & $5841 \pm 70$ & $0.23^{+0.06}_{-0.07}$ & $1.62 \pm 0.03$ & $4.6^{+1.6}_{-0.7}$&$5.0\pm 2.0$ & $16.5\pm 6.6$ & 17 & $5.0^{+0.4}_{-1.6}$  \\
TOI-1836 & $6198 \pm 100$ & $-0.1 \pm 0.1$ & $1.66 \pm 0.03$ & $3.5^{+1.3}_{-0.5}$&$5.9\pm 1.0$ & $14.3\pm 2.5$ & 18 & $4.8^{+0.9}_{-0.1}$  \\
TOI-2019& $5588 \pm 130$ & $0.4 \pm 0.1$ & $1.73 \pm 0.03$ &$6.0^{+1.0}_{-2.0}$ & $1.6\pm 1.0$ & $55.0\pm 34.4$ & 18 & $4.93^{+0.12}_{-0.11}$  \\
TOI-2134 & $4620 \pm 80$ & $0.13 \pm 0.04$ & $0.71^{+0.02}_{-0.03}$ & $3.8^{+5.5}_{-2.7}$&$0.78\pm 0.1$ & $46.6\pm 6.1$ & 19 & $7.1^{+2.3}_{-3.1}$  \\
TOI-2525 & $5096 \pm 110$ & $0.14 \pm 0.05$ & $0.78 \pm 0.04$ & $3.99^{+4.3}_{-2.6}$& $1.5\pm 0.3$ & $26.6\pm 5.4$ & 20 & $10.6^{+3.6}_{-2.2}$  \\
TOI-4994 & $5640 \pm 110$ & $0.16 \pm 0.09$ & $1.06 \pm 0.04$ & $6.3^{+4.2}_{-3.8}$& - & $46.2 \pm 3.1$ & 21 & $12.7^{+2.2}_{-0.8}$  \\
TOI-5678 & $5485 \pm 70$ & $0.0 \pm 0.01$ & $0.938 \pm 0.007$ & $8.5 \pm 3.0$ & $3.1\pm 1.0$ & $15.4\pm 5.0$ & 22 & $8.8^{+1.1}_{-1.0}$  \\
HD 332231 & $6089 \pm 100$ & $0.04 \pm 0.06$ & $1.28 \pm 0.04$ &$4.3^{+2.5}_{-1.9}$ & $5.3\pm 1.0$ & $12.3\pm 2.3$ & 23 & $3.0^{+0.5}_{-0.5}$  \\
PH2 & $5711 \pm 60$ & $-0.03 \pm 0.04$ & $0.961^{+0.016}_{-0.015}$ & - & $2.0\pm 1.0$ & $24.5\pm 12.2$ & 8 & $5.0 \pm 1.5$  \\
TOI-1898 & $6241 \pm 90$ & $-0.10 \pm 0.08$ & $1.62 \pm 0.03$& $3.4^{+1.0}_{-0.5}$ & $7.0\pm 1.0$ & $11.8\pm 1.7$ & 18 & $3.5^{+0.3}_{-1.1}$  \\
TOI-2328 & $5525 \pm 150$ & $0.09 \pm 0.03$ & $0.90\pm 0.06$ & $4.0^{+2.3}_{-2.7}$ & $2.61\pm 0.25$ & $17.6\pm 2.0$ & 24 & $7.3^{+1.5}_{-1.4}$  \\
\hline
NGTS-11 & $5050 \pm 80$ & $0.22 \pm 0.08$ & $0.832 \pm 0.013$ & $3.9 \pm 1.6$ & $1.1\pm 0.8$ & $38.5\pm 28.0$ & 25 & $0.6^{+0.5}_{-1.5}$  \\
Kepler-30 & $5498 \pm 60$ & $0.18 \pm 0.03$ & $0.95 \pm 0.12$ & - & $2.2\pm 1.0$ & $22\pm 10$ & 26,10 & $0.9^{+0.7}_{-1.6}$  \\
Kepler-539 & $5820 \pm 80$ & $-0.01 \pm 0.07$ & $0.952 \pm 0.034$ & $1.1^{+0.9}_{-0.1}$ & $4.4\pm 0.5$ & $11.0\pm 1.4$ & 27 & $0.9^{+0.5}_{-0.7}$  \\
Kepler-1656 & $5731 \pm 60$ & $0.19 \pm 0.04$ & $1.10 \pm 0.13$ & $6.31^{+2.1}_{-2.9}$&$2.8\pm 1.0$ & $20\pm 8$ & 28 & $1.6^{+0.9}_{-1.2}$  \\
K2-139 & $5340 \pm 110$ & $0.22 \pm 0.08$ & $0.92 \pm 0.04$ & $1.8 \pm 0.3$& $2.8\pm 0.6$ & $16.7\pm 3.6$ & 29 & $1.6^{+1.1}_{-1.8}$  \\
TOI-216 & $5054 \pm 110$ & $-0.16 \pm 0.09$ & $0.757 \pm 0.007$ & - & $5.7\pm 1.0$ & $6.8\pm 1.2$ & 30,31 & $0.42^{+0.14}_{-0.22}$  \\
TOI-1710 & $5665 \pm 55$ & $0.10 \pm 0.07$ & $0.968^{+0.016}_{-0.014}$ & $4.2^{+4.1}_{-2.7}$& $2.3\pm 0.4$ & $22\pm 4$ & 32 & $2.6^{+0.8}_{-2.1}$  \\
TOI-1823 & $4927 \pm 60$ & $0.24^{+0.1}_{-0.09}$ & $0.805^{+0.04}_{-0.03}$ & $6.3^{+6.3}_{-4.4}$& $4.0\pm 2.0$ & $10.2\pm 5.1$ & 17  & $1.8^{+1.5}_{-4.4}$  \\
TOI-2447 & $5730 \pm 80$ & $0.18 \pm 0.08$ & $1.006 \pm 0.009$ & $2 \pm 1$& $3.5\pm 0.6$ & $14.7\pm 2.5$ & 33 & $0.6^{+0.4}_{-1.0}$  \\
TOI-4010 & $4960 \pm 40$ & $0.37 \pm 0.07$ & $0.85 \pm 0.02$ &$6.3 \pm 3.1$& $1.3\pm 0.5$ & $33.3\pm 12.8$ & 34 & $1.5^{+1.2}_{-2.6}$  \\

    \hline
    \end{tabular}
    \begin{tablenotes}
    \item  Notes: The second reference for each star is for the $v\sin(i)$ value. $v\sin(i)$ values for TOI-1386, TOI-1439, and TOI-1823 were calculated from TRES spectra published on Exofop \citep{kurucz1992,tres,buchhave2010,buchhave2012}. For comparison, we list the ages calculated in the reference papers of the stellar parameters (if available). At the bottom of the table, we list the stars that did not satisfy the age requirement.
       \item References: 1 \cite{Chakraborty2018}; 2 \cite{Rainer2023}; 3 \cite{Nascimbeni2024}; 4 \cite{abu2022}; 5 \cite{Nowak2020}; 6 \cite{Vissapragada2020}; 7 \cite{Stassun2019}; 8 \cite{Dubber2019}; 9 \cite{Dalba2024}; 10 \cite{Petigura2022}; 11 \cite{Bruno2015}; 12 \cite{Yahalomi2024}; 13 \cite{Brewer2018}; 14 \cite{Borsato2019}; 15 (Dalba \& Peluso, private communication); 16 \cite{Hill2024}; 17 \cite{MacDougall2023}; 18 \cite{Chontos2024}; 19 \cite{Rescigno2024}; 20 \cite{Trifonov2023}; 21 \cite{martinez2025}; 22 \cite{Ulmer2023}; 23 \cite{Dalba2020}; 24 \cite{pinto2025}; 25 \cite{Gill2020}; 26 \cite{Fabrycky2012}; 27 \cite{Mancini2016}; 28 \cite{Brady2018}; 29 \cite{barragan2018}; 30 \cite{Dawson2019}; 31 \cite{Sharma2018}; 32 \cite{konig2022}; 33 \cite{Gill2024}; 34 \cite{Kunimoto2023}.
       \end{tablenotes}
     \end{threeparttable}}
\end{minipage}

\section{GASTLI results}
\begin{table*}
    \centering
    \resizebox{\textwidth}{!}{
    \begin{threeparttable}[b]
    \caption{\label{tab:planets} Properties and results of the interior structure modeling for the planets in our sample.}
    \begin{tabular}{lccccccccc}
    \hline
    \hline
      Planet   & Period~[d] & $R_p~[\text{R}_\oplus]$&$M_p~[\text{M}_\oplus]$&Age~[Gyr] & $T_{eq}$~[K]  & $f_{env,~low~Fe/H}$ & $f_{env,~medium~Fe/H}$ & $f_{env,~high~Fe/H}$ & References \\
      \hline

EPIC 211945201 b & $19.49213$ & $6.12 \pm 0.1$ & $27.0^{+14.0}_{-13.0}$ & $4.4^{+0.7}_{-1.0}$ & $882 \pm 15$ & $0.21 \pm 0.02$ & $0.29^{+0.06}_{-0.04}$ & $0.38 \pm 0.03$ & 1 \\
K2-24 b & $20.88257$ & $5.64 \pm 0.07$ & $20.6^{+1.6}_{-0.3}$ & $13.4^{+0.9}_{-0.4}$ & $769 \pm 37$ & $0.2 \pm 0.01$ & $0.27 \pm 0.03$ & $0.34 \pm 0.02$ & 2 \\
K2-24 c & $42.3773$ & $7.93^{+0.12}_{-0.13}$ & $16.4^{+1.3}_{-0.2}$ & $13.4^{+0.9}_{-0.4}$ & $607 \pm 30$ & $0.48 \pm 0.03$ & $0.59^{+0.07}_{-0.05}$ & $0.83^{+0.05}_{-0.07}$ & 2 \\
K2-280 b & $19.89526$ & $7.5 \pm 0.5$ & $37 \pm 6$ & $10.2^{+7.0}_{-1.3}$ & $788 \pm 26$ & $0.46 \pm 0.07$ & $0.57^{+0.11}_{-0.1}$ & $0.74 \pm 0.1$ & 3 \\
KOI-1783.01 & $134.4629$ & $9.03^{+0.23}_{-0.24}$ & $89.0^{+19.0}_{-16.0}$ & $3.4 \pm 0.7$ & $421 \pm 7$ & $0.72 \pm 0.05$ & $0.73^{+0.07}_{-0.06}$ & $0.93^{+0.04}_{-0.05}$ & 4 \\
KOI-1783.02 & $284.2162$ & $5.5^{+0.6}_{-0.3}$ & $18.5^{+5.5}_{-4.4}$ & $3.4 \pm 0.7$ & $328 \pm 6$ & $0.22^{+0.07}_{-0.05}$ & $0.28^{+0.09}_{-0.08}$ & $0.41^{+0.1}_{-0.09}$ & 4 \\
KOI-94 e & $54.3261$ & $6.07^{+0.12}_{-0.18}$ & $19 \pm 3$ & $2.5^{+0.4}_{-0.7}$ & $654 \pm 30$ & $0.2 \pm 0.02$ & $0.28^{+0.07}_{-0.05}$ & $0.39 \pm 0.04$ & 4 \\
Kepler-103 c & $179.60978$ & $5.45 \pm 0.18$ & $58 \pm 12$ & $5.2^{+0.5}_{-1.0}$ & $437 \pm 5$ & $0.15 \pm 0.03$ & $0.17^{+0.06}_{-0.04}$ & $0.27 \pm 0.05$ & 5 \\
Kepler-111 c & $224.77833$ & $7.08 \pm 0.33$ & $222.0^{+45.0}_{-42.0}$ & $2.6^{+0.9}_{-0.8}$ & $352 \pm 9$ & $0.27^{+0.07}_{-0.06}$ & $0.25^{+0.08}_{-0.07}$ & $0.56^{+0.11}_{-0.24}$ & 6 \\
Kepler-117 b & $18.7959228$ & $8.1 \pm 0.3$ & $30 \pm 11$ & $3.6^{+0.4}_{-0.3}$ & $990 \pm 30$ & $0.46 \pm 0.05$ & $0.58^{+0.09}_{-0.07}$ & $0.79^{+0.07}_{-0.08}$ & 7 \\
Kepler-1513 b & $160.8842$ & $8.6^{+0.04}_{-0.03}$ & $48.0^{+34.0}_{-20.0}$ & $4.1 \pm 1.9$ & $343 \pm 16$ & $0.69^{+0.07}_{-0.05}$ & $0.79^{+0.21}_{-0.08}$ & $0.98 \pm 0.02$ & 8 \\
Kepler-450 b & $28.4548844$ & $6.083 \pm 0.003$ & $19.4^{+11.1}_{-6.8}$ & $3.67^{+0.11}_{-0.17}$ & $863 \pm 35$ & $0.21 \pm 0.01$ & $0.27^{+0.07}_{-0.04}$ & $0.39^{+0.02}_{-0.01}$ & 9 \\
Kepler-849 b & $394.62508$ & $8.1 \pm 0.4$ & $299 \pm 70$ & $3.9^{+0.3}_{-0.23}$ & $363 \pm 12$ & $0.46 \pm 0.07$ & $0.45^{+0.11}_{-0.1}$ & $0.73^{+0.14}_{-0.23}$ & 6 \\
Kepler-9 b & $19.23891$ & $8.29^{+0.54}_{-0.43}$ & $43.4^{+1.6}_{-2.0}$ & $2.5^{+0.9}_{-1.1}$ & $724 \pm 11$ & $0.58^{+0.11}_{-0.08}$ & $0.69 \pm 0.1$ & $0.8^{+0.04}_{-0.06}$ & 10 \\
Kepler-9 c & $38.9853$ & $8.08^{+0.54}_{-0.41}$ & $29.9^{+1.1}_{-1.3}$ & $2.5^{+0.9}_{-1.1}$ & $572 \pm 9$ & $0.55^{+0.09}_{-0.08}$ & $0.68^{+0.12}_{-0.11}$ & $0.86^{+0.08}_{-0.1}$ & 10 \\
TIC 139270665 b & $23.624$ & $9.8 \pm 0.5$ & $164 \pm 17$ & $3.8 \pm 0.5$ & $834 \pm 22$ & $0.32 \pm 0.04$ & $0.33^{+0.07}_{-0.05}$ & $0.61^{+0.07}_{-0.05}$ & 11 \\
TOI-1386 b & $25.83839$ & $6.053^{+0.2}_{-0.18}$ & $47.0 \pm 6.3$ & $4.2^{+1.5}_{-1.2}$ & $676 \pm 14$ & $0.22 \pm 0.03$ & $0.27^{+0.07}_{-0.06}$ & $0.41^{+0.06}_{-0.05}$ & 12 \\
TOI-1439 b  & $27.644$ & $4.24^{+0.23}_{-0.37}$ & $38.5^{+5.7}_{-5.6}$ & $5.0^{+0.4}_{-1.6}$ & $819 \pm 13$ & $0.03^{+0.02}_{-0.01}$ & $0.03 \pm 0.02$ & $0.06 \pm 0.03$ & 13 \\
TOI-1836 b & $20.38085$ & $8.38 \pm 0.19$ & $28.4 \pm 4.3$ & $4.8^{+0.9}_{-0.1}$ & $999 \pm 21$ & $0.53 \pm 0.04$ & $0.66^{+0.08}_{-0.06}$ & $0.85^{+0.05}_{-0.07}$ & 14 \\
TOI-2019 b & $15.3444$ & $5.66 \pm 0.28$ & $34.6 \pm 4.2$ & $4.93^{+0.12}_{-0.11}$ & $989 \pm 29$ & $0.16 \pm 0.03$ & $0.22^{+0.06}_{-0.05}$ & $0.28^{+0.06}_{-0.07}$ & 14 \\
TOI-2134 c & $95.5$ & $7.27 \pm 0.42$ & $42 \pm 8$ & $7.1^{+2.3}_{-3.1}$ & $308 \pm 7$ & $0.47^{+0.08}_{-0.07}$ & $0.55^{+0.1}_{-0.09}$ & $0.78^{+0.11}_{-0.12}$ & 15 \\
TOI-2525 b & $23.2856$ & $8.67 \pm 0.12$ & $27 \pm 2$ & $10.6^{+3.6}_{-2.2}$ & $560 \pm 16$ & $0.67 \pm 0.03$ & $0.79^{+0.11}_{-0.06}$ & $0.98^{+0.01}_{-0.02}$ & 16 \\
TOI-4994 b & $21.491984$ & $8.54^{+0.34}_{-0.31}$ & $89 \pm 13$ & $12.7^{+2.2}_{-0.8}$ & $718 \pm 21$ & $0.63 \pm 0.06$ & $0.73^{+0.09}_{-0.07}$ & $0.94^{+0.04}_{-0.06}$ & 17 \\
TOI-5678 b & $47.73022$ & $4.91 \pm 0.08$ & $20.0 \pm 4.0$ & $8.8^{+1.1}_{-1.0}$ & $513 \pm 7$ & $0.13 \pm 0.01$ & $0.16 \pm 0.02$ & $0.23 \pm 0.02$ & 18 \\
HD 332231 b & $18.71204$ & $9.7^{+0.3}_{-0.28}$ & $77.6 \pm 6.7$ & $3.0 \pm 0.5$ & $876 \pm 22$ & $0.78^{+0.06}_{-0.05}$ & $0.89 \pm 0.07$ & $0.98^{+0.02}_{-0.03}$ & 19 \\
PH2 b & $282.5254$ & $9.49 \pm 0.16$ & $109.0^{+30.0}_{-33.0}$ & $5.0 \pm 1.5$ & $295 \pm 5$ & $0.83^{+0.05}_{-0.05}$ & $0.9^{+0.06}_{-0.06}$ & $0.99^{+0.01}_{-0.01}$ & 5 \\
TOI-1898 b & $45.522149$ & $9.74 \pm 0.22$ & $127.5 \pm 6.6$ & $3.5{+0.3}_{-1.1}$ & $741 \pm 13$ & $0.77 \pm 0.04$ & $0.84^{+0.05}_{-0.04}$ & $0.99^{+0.01}_{-0.02}$ & 14 \\
TOI-2328 b & $17.10197$ & $9.97 \pm 0.45$ & $51.0 \pm 6.4$ & $7.3^{+1.5}_{-1.4}$ & $707 \pm 31$ & $0.87 \pm 0.08$ & $0.92^{+0.05}_{-0.08}$ & $0.98^{+0.01}_{-0.03}$ & 20 \\

    \hline
    \end{tabular}
        \begin{tablenotes}
       \item References: 1 \cite{Chakraborty2018}; 2 \cite{Nascimbeni2024}; 3 \cite{Nowak2020}; 4 \cite{Jontof2022}; 5 \cite{Dubber2019}; 6 \cite{Dalba2024}; 7 \cite{Bruno2015}; 8 \cite{Yahalomi2024}; 9 \cite{Yoffe2021}; 10 \cite{Borsato2019}; 11 (Dalba \& Peluso, private communication); 12 \cite{Hill2024}; 13 \cite{Polanski2024}; 14 \cite{Chontos2024}; 15 \cite{Rescigno2024}; 16 \cite{Trifonov2023}; 17 \cite{martinez2025}; 18 \cite{Ulmer2023}; 19 \cite{Dalba2020}; 20 \cite{pinto2025}.
       \end{tablenotes}
     \end{threeparttable}}
\end{table*}
\end{appendix}

\end{document}